\documentclass[12pt]{article}
\usepackage{amssymb,amsfonts,amsthm}
\usepackage{amsmath}
\usepackage{bm}             
\usepackage[mathscr]{eucal}
\usepackage{cancel}
\usepackage{wasysym}
\usepackage{color}
\usepackage{graphicx}
\usepackage{subfigure}
\def\be{\begin{equation}}
\def\ee{\end{equation}}
\def\bea{\begin{eqnarray}}
\def\eea{\end{eqnarray}}

\def\hsp5{\hspace{5mm}}

\DeclareMathOperator{\arctanh}{arctanh}

\theoremstyle{remark}

\newcommand{\sfrac}[2]{{\textstyle{#1\over#2}}}

\title{\sc Tracking Quintessence}

\begin{document}

\author{
\sc Artur Alho,$^{1}$\thanks{Electronic address:{\tt
aalho@math.ist.utl.pt}}\,, Claes Uggla,$^{2}$\thanks{Electronic address:{\tt
claes.uggla@kau.se}}\,  and John Wainwright$^{3}$\thanks{Electronic
address:{\tt jwainwri@uwaterloo.ca}}\\
$^{1}${\small\em Center for Mathematical Analysis, Geometry and Dynamical Systems,}\\
{\small\em Instituto Superior T\'ecnico, Universidade de Lisboa,}\\
{\small\em Av. Rovisco Pais, 1049-001 Lisboa, Portugal.}\\
$^{2}${\small\em Department of Physics, Karlstad University,}\\
{\small\em S-65188 Karlstad, Sweden.}\\
$^{3}${\small\em Department of Applied Mathematics, University of Waterloo,}\\
{\small\em Waterloo, ON, N2L 3G1, Canada.}}


\date{}
\maketitle

\begin{abstract}

Tracking quintessence, in a spatially flat and isotropic space-time with a
minimally coupled canonical scalar field and an asymptotically inverse power-law
potential $V(\varphi)\propto\varphi^{-p}$, $p>0$, as $\varphi\rightarrow0$,
is investigated. This is done by introducing a new
three-dimensional \emph{regular} dynamical system, which enables
a rigorous explanation of the tracking
feature: 1) The dynamical system has a tracker fixed point $\mathrm{T}$ with a
two-dimensional stable manifold that pushes an open set of nearby solutions
toward a single 
tracker solution originating from $\mathrm{T}$.
2) All solutions, including the tracker solution and the solutions that
track/shadow it, end at a common future attractor fixed point that
depends on the potential. Thus, the open set of solutions that shadow the
tracker solution share its properties during the tracking quintessence epoch.
We also discuss similarities and differences of underlying mechanisms
for tracking, thawing and scaling freezing quintessence, and, moreover,
we illustrate with state space pictures that all of these types of
quintessence exist simultaneously for certain potentials.


\end{abstract}

\section{Introduction}

In 1998 observations of type Ia supernovae indicated that the
Universe is presently accelerating~\cite{rieetal98,peretal99}.
Within the framework of General Relativity this cosmic acceleration implies that
there exists an exotic energy component in the Universe, called dark energy,
with an equation of state parameter satisfying $w_\mathrm{DE} < -1/3$.
The simplest candidate for dark energy, apart from a cosmological constant,
is a dynamical canonical scalar field $\varphi$, minimally coupled to
gravity and with a potential $V(\varphi)$, referred to as quintessence,
Caldwell {\it et al}. (1998)~\cite{caletal98} (for reviews and references
about quintessence, see {\it e.g.} Tsujikawa (2013)~\cite{tsu13} and Bahamonde
{\it et al.} (2018)~\cite{bahetal18}). At the present time the energy density of quintessence
and matter are roughly of the same size ($\Omega_\varphi\approx 0.7$, $\Omega_\mathrm{m}\approx 0.3$)
which raises the need to explain this near coincidence without specifying precise
initial conditions. To address this difficulty Zlatev {\it et al}. (1999)~\cite{zlaetal99},
and Steinhardt {\it et al}. (1999)~\cite{steetal99} showed that for potentials with the
property $V(\varphi)\propto\varphi^{-p}$, $p>0$, as $\varphi\rightarrow0$, the quintessence
governing equations have a special solution they called a tracker solution since it attracts
solutions that then track/shadow it for a wide range of initial conditions.
A second property the tracker solution
exhibit is that the scalar field equation of state parameter $w_\varphi$ is nearly
constant in the matter-dominated epoch and less than $w_\mathrm{m}$, the equation
of state parameter of the background matter. This implies
that the energy density $\rho_\mathrm{m}$ decreases
faster than $\rho_\varphi$ so that as the universe evolves from
matter-domination $\rho_\varphi$ will catch up and overtake $\rho_\mathrm{m}$. Taken together these
two properties go some way  toward solving the coincidence problem. This type of evolution is
referred to as \emph{tracking quintessence}.\footnote{We
note that prior to the development of the concept
of tracker quintessence Peebles and Ratra (1988)~\cite{peerat88} and~\cite{ratpee88} studied
scalar field cosmologies with matter and showed that for the inverse power-law potential when
$\rho_\varphi\ll\rho_{\mathrm m}$ one obtains $\rho_\varphi/\rho_{\mathrm m}\sim e^{6/(2+p)N}$,
which shows that $\rho_\varphi$ decreases more slowly than $\rho_{\mathrm m}$ as the
universe expands and will eventually become dominant. This is consistent
with the `tracker' expression $w_\varphi := p_\varphi/\rho_\varphi = -2/(2+p)$, first
derived by Ratra and Quillen (1992)~\cite{RatQui92}, eq. (5), and then by Steinhardt {\it et al}.
(1999)~\cite{steetal99} for the inverse power-law potential, see also Podario and
Ratra (2000)~\cite{PodRat00} and Peebles and Ratra (2003)~\cite{peerat03}.}

In this paper we give a description of tracking quintessence by means of a new \emph{regular}
dynamical systems framework that is valid for asymptotically inverse power-law potentials,
$V(\varphi)\propto\varphi^{-p}$, $p>0$, as $\varphi\rightarrow0$.\footnote{This
paper complements our recent paper Alho, Uggla and Wainwright (2023)~\cite{alhetal23} which deals
with potentials for which $\lambda(\varphi) = -V_{,\varphi}/V$ is bounded and $\varphi \in (-\infty,\infty)$,
referred to as AUW~\cite{alhetal23}. This is in contrast both physically and mathematically to
the present asymptotically inverse power-law potentials, which yield an extremely steep
potential wall when $\varphi \rightarrow 0$, resulting in $\varphi \in (0,\infty)$
instead of $\varphi \in (-\infty, \infty)$.}
We use our new regular dynamical systems framework to show that there exists a
unique tracker solution with the properties of the heuristically defined tracker
solution of Steinhardt and co-workers, originating from a new \emph{hyperbolic tracker
fixed point}. The fact that all the eigenvalues of this isolated fixed point have 
non-zero real parts makes it possible to i) explain the tracking feature and ii) obtain analytic 
approximate expressions for tracker solutions. Moreover, the global and regular structure of 
the state space shows explicitly a) the entire tracker solution and gives insight into the possible 
initial conditions which lead to solutions that approach the tracker fixed point and then track/shadow
the tracker solution; b) that these models also give rise to thawing quintessence solutions,
and that some potentials belonging to the present class simultaneously in addition give rise
to co-existing scaling freezing quintessence solutions.

The outline of the paper  is as follows. In the next section we derive the
new regular dynamical system. In Section~\ref{prop.dyn.sys} we briefly review
the tracker conditions of Steinhardt and co-workers~\cite{steetal99,zlaetal99} and
describe the general structure of the new state space, including the fixed points of the
dynamical system that determine the intermediate and late time behaviour of the quintessence
solutions. In Section~\ref{sec:cosmconsequences} we give a general dynamical systems
description of thawing, scaling freezing and tracking quintessence, focussing on
the latter, and in Section~\ref{sec:graphs} we give specific examples that illustrate the
previous general discussion using three-dimensional state space
figures and graphs of $w_\varphi$. Finally, there is a brief concluding Section~\ref{sec:concl}.

\section{A regular dynamical system for unbounded $\lambda$  \label{deriv.dyn.sys}}

Consider a spatially flat and isotropic Robertson-Walker spacetime with a
source that consists of matter with an energy density $\rho_\mathrm{m}>0$ and
pressure $p_\mathrm{m}=0$, which represents cold dark matter,\footnote{This
simple model is useful for describing the transition from
an epoch of matter-domination to an epoch in which the scalar field is dominant.
A more realistic model requires a two component source with
matter and radiation leading to a four-dimensional state space,
as described in~\cite{alhugg23}.}
and a minimally coupled canonical scalar field, $\varphi$, with a
potential $V(\varphi)>0$, which results in the following energy density and pressure
\begin{equation}\label{rhophipphi}
\rho_\varphi = \frac12\dot{\varphi}^2 + V(\varphi),\qquad
p_\varphi = \frac12\dot{\varphi}^2 - V(\varphi),
\end{equation}
where an overdot represents the derivative with
proper time $t$. For these models, the Raychaudhuri equation,
the Friedmann equation, the non-linear Klein-Gordon
equation, and the energy conservation law for matter with zero pressure,
can be written as\footnote{We use units such
that $c=1$ and $8\pi G=1$, where $c$ is the speed of light and
$G$ is the gravitational constant.}
\begin{subequations}\label{Mainsysdim}
\begin{align}
\dot{H} + H^2 &= -\frac16(\rho + 3p), \label{Ray}\\
3H^2 &= \rho, \label{Gauss}\\
\ddot{\varphi} &=-3H\dot{\varphi} - V_{,\varphi}, \label{KG}\\
\dot{\rho}_\mathrm{m} &= -3H\rho_\mathrm{m}, \label{matter.evol}
\end{align}
\end{subequations}
where the Hubble variable is defined by $H=\dot a/a$, and the total energy density
$\rho$ and pressure $p$ are given by
\begin{equation}\label{rho.p}
\rho = \rho_\varphi + \rho_\mathrm{m},\qquad p = p_\varphi.
\end{equation}

Our dynamical system is based on three key quantities associated with the
scalar field: the scalar field equation of state parameter
$w_\varphi := p_\varphi/\rho_\varphi$, the Hubble-normalized
energy density $\Omega_\varphi$ and scalar field potential $\Omega_V$:
\begin{subequations} \label{var.scalar}
\begin{align}
1+w_\varphi &:= \frac{\rho_\varphi +p_\varphi}{\rho_\varphi} =
\frac{\varphi'^2 }{3\Omega_\varphi},\quad
\text{provided that}\quad \Omega_\varphi>0, \label{w.varphi}\\
\Omega_\varphi &:= \frac{\rho_\varphi}{3H^2} = \frac16\varphi'^2 + \Omega_V,\\
\Omega_V &:= \frac{V}{3H^2}=\frac12(1-w_\varphi)\,\Omega_\varphi,\label{OmVdef}
\end{align}
\end{subequations}
where a ${}^\prime$ denotes the derivative with respect to $e$-fold time,
$N := \ln(a/a_0)$, where $a$ is the cosmological scale factor and $a_0 = a(t_0)$
is its value at the present time, given by $ N=0$.
It follows from~\eqref{rhophipphi} and~\eqref{var.scalar} and the
assumed positivity of $V$ that
\begin{equation}
1+w_{\varphi} \geq 0,\qquad 1-w_{\varphi} > 0.
\end{equation}

When using equations~\eqref{Mainsysdim} we will convert from proper time
$t$ to $e$-fold time $N$ using
\begin{equation} \label{t.N}
\partial_t=H\partial_N, \quad \partial_t^2=H^2[\partial_N^2 -(1+q)\partial_N],
\end{equation}
where the \emph{deceleration parameter} $q$ is defined by
\begin{equation} \label{q.def}
1+q := -\frac{\dot H}{H^2} .
\end{equation}
In particular~\eqref{KG} assumes the form
\begin{equation} \label{KG.N}
\varphi''+(2-q)\varphi'-3\lambda(\varphi)\Omega_V=0,
\end{equation}
where $\lambda(\varphi)$ is defined by
\begin{equation} \label{lambda.def}
\lambda := - \frac{V_{,\varphi}}{V}.
\end{equation}
We also need the Hubble-normalized matter density $\Omega_\mathrm{m}$
and its evolution equation, given by
\begin{equation}
\Omega_\mathrm{m} := \frac{\rho_\mathrm{m}}{3H^2}=1-\Omega_\varphi,\qquad
\Omega_\mathrm{m}'= (2q-1)\Omega_\mathrm{m},\label{Ommdef}
\end{equation}
as follows from~\eqref{matter.evol} and~\eqref{q.def}.
The deceleration parameter $q$ can be expressed as
\begin{equation}  \label{q.expr}
1+q = \frac32\left(1 + w_\varphi\Omega_\varphi\right),
\end{equation}
as follows from~\eqref{q.def} in conjunction with~\eqref{rhophipphi}, \eqref{Mainsysdim}
and~\eqref{rho.p}.


Equation~\eqref{w.varphi}, which can be written as
\begin{equation}\label{w.varphi.1}
(\varphi')^2 = 3\Omega_\varphi(1+w_\varphi),
\end{equation}
relates $\varphi'$ to $\Omega_\varphi$ and $w_\varphi$,
which satisfy the differential equations\footnote{Equation~\eqref{Om.prime}
follows from~\eqref{Ommdef} and~\eqref{q.def}; equation~\eqref{w.var.prime} is obtained by
differentiating~\eqref{w.varphi.1} and using~\eqref{Om.prime} and~\eqref{KG.N}.}
\begin{subequations}\label{alt.evol.eqs}
\begin{align}
\Omega_\varphi' &= -3w_\varphi(1-\Omega_\varphi)\Omega_\varphi,\label{Om.prime}\\
w_\varphi' &= -3(1- w_\varphi)
\left(1+w_\varphi-\frac{1}{3}\lambda(\varphi)\varphi'\right).\label{w.var.prime}
\end{align}
\end{subequations}
As in AUW~\cite{alhetal23}, since $1+w_\varphi \geq 0$ we can replace
$w_\varphi$ by a variable $u$ according to
\begin{equation}\label{def.u}
u^2 := 1+w_\varphi,
\end{equation}
with the stipulation that $u$ has the \emph{same sign} as $\varphi'$. It follows
from~\eqref{w.varphi.1} that
\begin{equation}\label{varphi.prime}
\varphi'= \sqrt{3\Omega_\varphi}\,u,
\end{equation}
which leads to that~\eqref{w.var.prime} takes the form
\begin{equation}\label{u.prime0}
u^\prime = \frac32(2-u^2)\left(\lambda(\varphi)\sqrt{\Omega_\varphi/3} - u\right).
\end{equation}

In this paper we consider positive asymptotically inverse power-law potentials
for which $\lambda$ has the following divergence as $\varphi\rightarrow 0$:
\begin{subequations} \label{lambda.prop}
\begin{equation} \label{lambda.lim.0}
\lim_{\varphi\rightarrow 0}\varphi\lambda= p,
\end{equation}
with $p>0$ where $\lambda(\varphi)$ is subsequently assumed to be bounded
with a finite limit as $\varphi\rightarrow \infty$,
\begin{equation} \label{lambda.lim.infty}
\lim_{\varphi\rightarrow\infty}\lambda = \lambda_+.
\end{equation}
\end{subequations}

Next we replace the unbounded scalar field variable $\varphi\in [0,\infty)$ by
a bounded variable $\bar\varphi\in [0,1]$. We do this by choosing a
regular, increasing (and hence invertible) function $\bar{\varphi}(\varphi)$ with
$\varphi\in [0,\infty)$. The choice of $\bar{\varphi}(\varphi)$ is guided by the
form of $\lambda(\varphi)$, but is required to satisfy the following
conditions:
\begin{equation} \label{bar.phi}
\bar{\varphi}(0)=0, \qquad \left.\frac{d\bar{\varphi}}{d\varphi}\right|_{\varphi=0} = b>0, \qquad
\lim_{\varphi\rightarrow\infty} \bar{\varphi}(\varphi) = 1, \qquad
\lim_{\varphi\rightarrow\infty}\left(\frac{d\bar{\varphi}}{d\varphi}\right) = 0.
\end{equation}
We will regard the derivative $d\bar{\varphi}/d\varphi$ as a function of $\bar{\varphi}$
which we denote by $F(\bar{\varphi})$:
\begin{equation}\label{def.F}
F(\bar{\varphi}) := \frac{d\bar{\varphi}}{d\varphi},\quad\text{with}\quad F(0)=b,\quad F(1)=0,
\end{equation}
where the two equalities follow from~\eqref{bar.phi}.
Hence~\eqref{varphi.prime} assumes the form
\begin{equation}\label{bar.var.prime}
\bar{\varphi}' = \sqrt{3\Omega_\varphi}u\,F(\bar{\varphi}).
\end{equation}

We now come to the main new ingredient in our new dynamical systems formulation
where we use the growth condition~\eqref{lambda.lim.0}
to regularize equations~\eqref{u.prime0} and~\eqref{bar.var.prime}.
It follows from~\eqref{lambda.lim.0} that\footnote{Use
$\lim_{\bar\varphi\rightarrow 0}(\bar\varphi/\varphi)=b$,
which follows from the second equation in~\eqref{bar.phi}.}
$\lim_{\bar\varphi\rightarrow 0}(\bar\varphi\lambda(\bar\varphi)) = p\,b$,
where, with a slight abuse of notation, $\lambda(\bar\varphi)=\lambda(\varphi(\bar{\varphi}))$.
This makes it possible to define a regular function $G(\bar\varphi)$ according to
\begin{subequations}
\begin{equation}\label{G.def}
G(\bar\varphi) := \bar\varphi\,\lambda(\bar\varphi),
\quad\text{for}\quad 0<\bar\varphi\leq1,\quad G(0)=p\,b,
\end{equation}
while~\eqref{lambda.lim.infty} and~\eqref{bar.phi}
leads to $\lambda|_{\bar{\varphi}=1}=\lambda_{+}$ and hence that
\begin{equation} \label{prop.G}
G(1)=\lambda_{+}.
\end{equation}
\end{subequations}
We then use~\eqref{G.def} to replace $\lambda$ by $G$
in~\eqref{u.prime0}, which suggests that we
define a new \emph{positive} variable $v$ by writing
\begin{equation} \label{Om.v}
\Omega_\varphi=3v^2\bar\varphi^2,\qquad v := \frac{1}{\bar{\varphi}}\sqrt{\frac{\Omega_\varphi}{3}}.
\end{equation}
After substituting the above expression for $\Omega_\varphi$ in~\eqref{u.prime0}
and~\eqref{bar.var.prime} we obtain regular equations for $u'$ and $\bar\varphi'$.
The final step is to differentiate~\eqref{Om.v} and use~\eqref{Om.prime}
and~\eqref{bar.var.prime} to calculate $v'$. On collecting the results, the regular
system of equations for the state vector $(\bar\varphi,u,v)$ has the following form:
\begin{subequations}\label{Dynsysuv}
\begin{align}\
\bar{\varphi}^\prime &=3uv\bar{\varphi}F(\bar{\varphi}),\label{phi.prime}\\
u^\prime &= \frac32(2-u^2)(vG(\bar{\varphi}) - u),\label{u.prime}\\
v^\prime &= \frac32\left[(1 - u^2)(1 - 3v^2\bar{\varphi}^2) -
2uv F(\bar{\varphi})\right]v,
\end{align}
\end{subequations}
where $F(\bar\varphi)$ and $G(\bar\varphi)$ are defined by~\eqref{def.F}
and~\eqref{G.def}.

We conclude this section by noticing that tracking quintessence was discovered
in connection with the inverse power-law potential $V \propto \varphi^{-p}$, $p>0$.
To treat these models in the present dynamical systems setting we can
follow~\cite{alhugg15b} and define $\bar{\varphi}$ as
\begin{equation}
\bar{\varphi} := \frac{\varphi}{1 + \varphi},
\end{equation}
which results in a regular dynamical system with
\begin{equation}
F = (1 - \bar{\varphi})^2,\qquad G = p(1 - \bar{\varphi}),
\end{equation}
from which it follows that $F(0)=1$, and thereby $b=1$,
$G(0) = p$, $G(1) = 0$.

\section{Tracker conditions and general dynamical systems features\label{prop.dyn.sys}}

\subsection{Tracker conditions}\label{sec:trackcond}

Steinhardt and co-workers~\cite{steetal99,zlaetal99} introduced the concept of
\emph{tracking quintessence}, which entailed that a wide range of initial conditions
result in solutions that are attracted to a special solution called the tracker
solution. The analysis in Steinhardt {\it et al}.~\cite{steetal99}
uses the scalars $\Delta$ and $\Gamma$, defined by:
\begin{equation}
\Delta := \pm\lambda\,\sqrt{\frac{\Omega_\varphi}{3(1+w_\varphi)}},\qquad
\Gamma := \frac{VV_{,\varphi\varphi}}{V_{,\varphi}^2} = 1 + (\lambda^{-1})_{,\varphi}
\end{equation}
(for $\Delta$, see equation (9) in~\cite{steetal99}).
The definition of $\Delta$ results in that the evolution equation for $w_\varphi$
can be written in the form
\begin{equation}
w_\varphi'=-3(1-w_\varphi^2)(1-\Delta),
\end{equation}
while \emph{if} $w_\varphi'=0$ and $1+w_\varphi>0$ then\footnote{See
equation (33) in Rubano {\it et al}. (2004)~\cite{rubetal04}.}
\begin{equation} \label{gamma.constr}
\Gamma - 1 =\frac{(w_\mathrm{m}-w_\varphi)(1-\Omega_\varphi)}{2(1+w_\varphi)}.
\end{equation}

In the approach of Steinhart {\it et al}.~\cite{steetal99} tracking quintessence is
described by the conditions
\begin{equation}\label{trackercond}
\Delta \approx 1 \quad \text{and}\quad \Gamma - 1 > 0.
\end{equation}
Heuristically, the first condition implies that $w_\varphi$ is nearly constant, and hence
that~\eqref{gamma.constr} holds approximately, which, due to the second tracker condition,
$\Gamma-1>0$, suggests that $w_\varphi<w_\mathrm{m}$, where $w_\mathrm{m} = 0$ in the present paper
results in $w_\varphi<0$. This condition is in turn important since it implies that, as long
as the conditions hold, $\Omega^{\prime}_\varphi>0$ according to~\eqref{Om.prime}
so that $\Omega_\varphi$ increases as the universe ages. Finally, the second condition,
$\Gamma - 1 = (\lambda^{-1})_{,\varphi} > 0$, corresponds to that $\lambda$ is
monotonically decreasing.
In terms of our state space variables the scalar $\Delta$ becomes
\begin{equation}\label{DeltaG}
\Delta = \bar{\varphi}\lambda(\bar{\varphi})\left(\frac{v}{u}\right) = G(\bar{\varphi})\left(\frac{v}{u}\right).
\end{equation}
%

\subsection{State space features}

Recall that the dynamical system~\eqref{Dynsysuv} asymptotically depends on the three
parameters $b=F(0)$, $\lambda_{+}=G(1)$ and $p$ via $G(0) = p\,b$, where $b$ is
associated with the bounded scalar field variable $\bar{\varphi}$, while $\lambda_+$ and $p$
characterize the asymptotic properties of the scalar field potential $V(\bar{\varphi})$.
The state space of the system~\eqref{Dynsysuv} is described
by the bounded variables $\bar{\varphi}\in [0,1]$, $u \in [-\sqrt{2},\sqrt{2}]$,
and the unbounded variable $v\in[0,\infty)$. There are six invariant boundary sets:
\begin{equation} \label{boundary}
\bar{\varphi} = 0,\qquad \bar{\varphi} = 1,\qquad u = \pm\sqrt{2},\qquad v=0,
\qquad v = \frac{1}{\sqrt{3}\,\bar{\varphi}}.
\end{equation}
Thinking of $(\bar\varphi,u,v)$ as Cartesian coordinates the state space
can be visualized as a  three-dimensional solid, with rectangular base
$v=0$, vertical sides $u =\pm\sqrt{2}$, and vertical ends $\bar\varphi=0,1$. The solid
is bounded above by the curved `ski-slope surface' $v=1/\sqrt{3}\bar{\varphi}$,
$0<\bar{\varphi}\leq 1$, with $v\rightarrow\infty$ as $\bar{\varphi}\rightarrow 0$.
We will refer to this three-dimensional solid as the `ski-slope state space'.

The relation~\eqref{Om.v}, $\Omega_\varphi = 3v^2\bar{\varphi}^2$,
provides a physical interpretation of some of the boundary sets. First, on the sets $v=0$
and $\bar{\varphi}=0$ we have $\Omega_\varphi=0$, and hence $\Omega_\mathrm{m}=1$.
Thus the subset on which the \emph{matter is dominant} $(\Omega_m=1)$ is the union of the
boundary sets $v=0$ and $\bar{\varphi}=0$. Second, the ski-slope surface $v=1/\sqrt{3}\bar{\varphi}$
is the boundary set on which the \emph{scalar field is dominant}, since $\Omega_\varphi=1$.
To continue, it follows from~\eqref{def.u} that the sets $u=\pm\sqrt2$ are characterized by
$w_\varphi=1$ and hence, as follows from~\eqref{OmVdef}, $\Omega_V=0$.
Finally, recall that $\lambda$ has a constant value
$\lambda_+$ on the invariant set $\bar{\varphi}=1$, which is therefore referred to as
\emph{the exponential potential boundary set}.

\begin{figure}[ht!]
	\begin{center}
	\subfigure[The ski-slope state space with the six boundary sets in~\eqref{boundary}.
     Depicted are also solution trajectories on the $v=0$ subset with $\bar{\varphi} = \mathrm{constant}$,
     which are independent of the potential, as are
     the solution trajectories on the $u=\pm \sqrt{2}$ boundaries.]{\label{Fig1a}
	\includegraphics[width=0.45\textwidth]{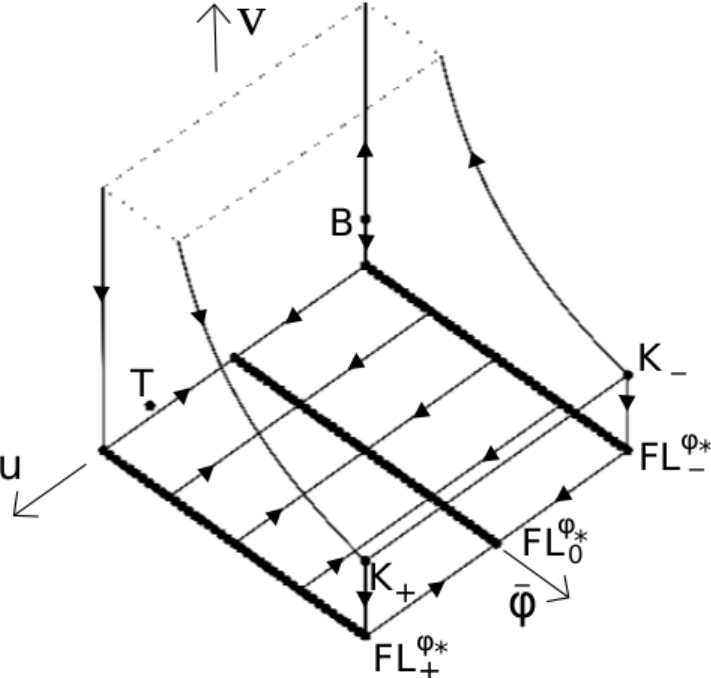}}
\hspace{0.5cm}
		\subfigure[The solution space structure on the $\bar{\varphi}=0$ boundary, where there are
         no bifurcations as $b>0$ and $p>0$ change.
         The dashed curves show where
         $v^\prime=0$; $v$ is increasing (decreasing) between (outside) the two $v^\prime=0$ curves.
         \emph{On} this boundary the fixed point $\mathrm{B}$ is a
         source while $\mathrm{T}$ is a sink.
]{\label{Fig1b}
		\includegraphics[width=0.4\textwidth]{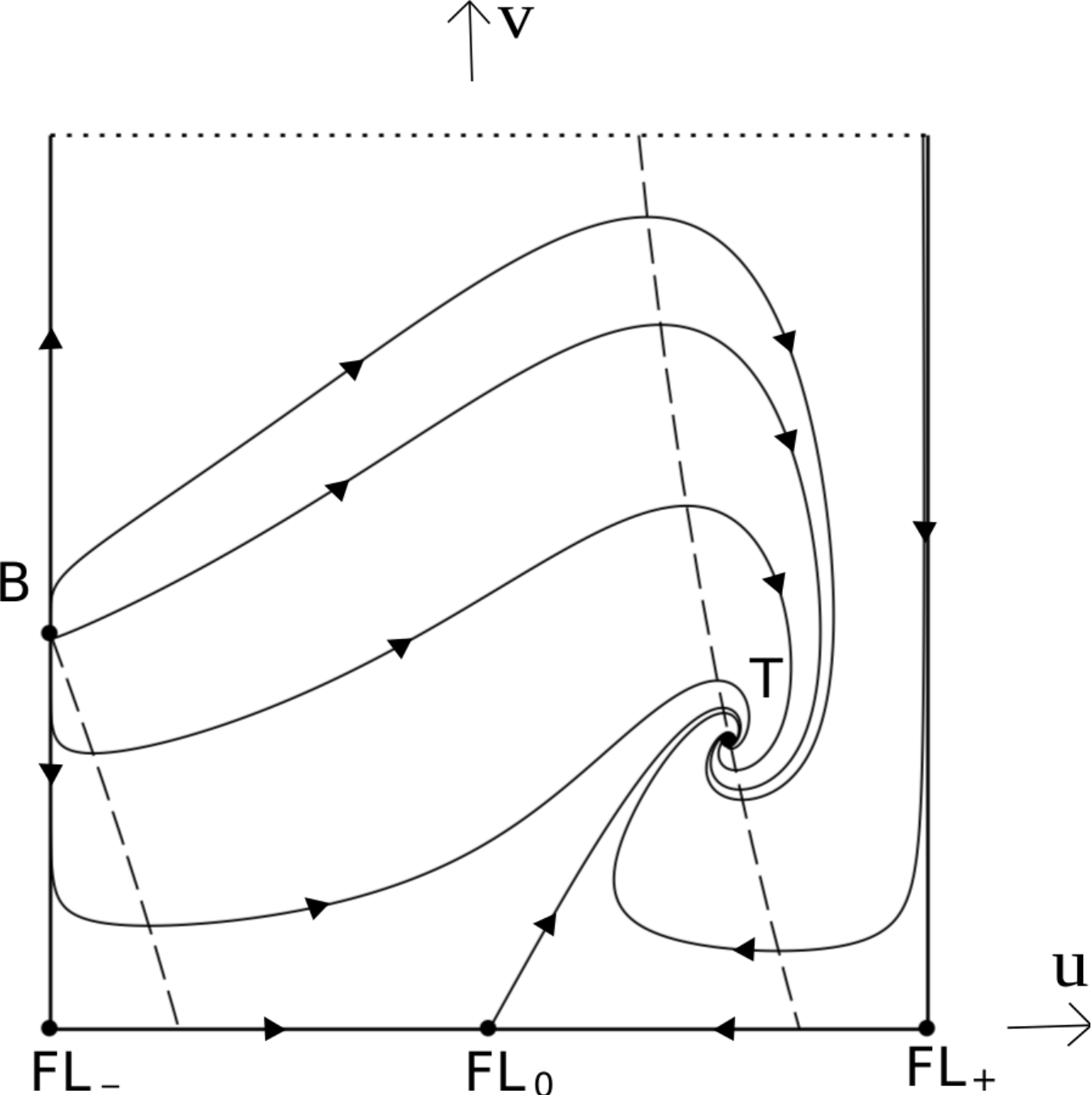}}
		\vspace{-0.5cm}
	\end{center}
	\caption{The ski-slope state space and the $\bar{\varphi}=0$ boundary.
             The dotted lines correspond to a cut off in $v$ in order to obtain
             a finite figure, since $v\rightarrow \infty$ when
             $\bar\varphi\rightarrow 0$, which also results in a
             cut off for small $\bar{\varphi}$ on the scalar field dominant
             boundary $v\bar{\varphi} = 1/\sqrt{3}$.}\label{Fig1}
\end{figure}

We note that there is a mathematical common ground between the present paper
and AUW~\cite{alhetal23}. In particular, the boundary sets $v=0$, on which
$\bar{\varphi} = \mathrm{constant}$, and $\bar{\varphi}=1$
are \emph{identical} to the same boundary sets in
AUW~\cite{alhetal23}.\footnote{This can be seen by comparing the current
system~\eqref{Dynsysuv} with the corresponding system given by equations
(20) in AUW~\cite{alhetal23}, but note that the domain of $\bar{\varphi}$
(and $\varphi$) is different. Although $u$ is the same in AUW as here,
this is not the case with $v$. Since $v$ in AUW~\cite{alhetal23} is given
by $v(\text{AUW}) = \sqrt{\Omega_\varphi/3}$ it follows that the present $v$ and
$v(\text{AUW})$ are related by $v = v(\text{AUW})/\bar{\varphi}$. Thus $v = v(\text{AUW})$
when $\bar{\varphi}=1$, but note that it is the factor $\bar{\varphi}^{-1}$
in $v = v(\text{AUW})/\bar{\varphi} \rightarrow v(\text{AUW})/\varphi$ when
$\varphi\rightarrow 0$ that regularizes the present dynamical system
at $\varphi=0$, which enables our results concerning the tracker solution.}
The solution space structures on the exponential boundary arising
from different values of $\lambda_+$ were dealt with in detail in AUW and in~\cite{alhetal22},
and we will therefore not discuss them in this paper.
%
Figure~\ref{Fig1} illustrates some key features of the ski-slope state space; note that the
solution space structure on the boundaries $v=0$ and $u = \pm \sqrt{2}$ is independent of
$\lambda$ and thereby on the potential.

Since the scalar field potential is not exponential, the scalar field contributes
a source that is not scale-invariant. From this it follows that the Hubble scalar $H$ is
a function of the state vector $(\bar{\varphi},u,v)$. By using~\eqref{OmVdef}
in conjunction with~\eqref{def.u} and~\eqref{Om.v} we obtain
\begin{equation}
3H^2=\frac{V}{\Omega_V}= \frac{2V}{(1-w_{\varphi})\Omega_{\varphi}}
 = \frac{2V(\bar{\varphi})}{3(2-u^2)v^2\bar{\varphi}^2},
\end{equation}
where
\begin{equation}\label{3H2mon}
(3H^2)^\prime = - 2(1+q)(3H^2) = -3(1 + w_\varphi\Omega_\varphi)(3H^2) = -3[1 + 3(u^2-1)v^2\bar{\varphi}^2](3H^2),
\end{equation}
as follows from the definitions of the new variables and~\eqref{q.expr}.
In the interior state space, where $\Omega_\varphi<1$ and $-1 \leq w_\varphi < 1$,
it follows that $1+q>0$ and hence that $3H^2(u,v,\bar{\varphi})$
is monotonically decreasing. Thus,
there are no periodic orbits ({\it i.e.} solution trajectories)
or fixed points in the interior state space and
hence all fixed points of the dynamical system~\eqref{Dynsysuv} lie on
the boundary of the ski-slope state space, given by~\eqref{boundary}
(in addition, an asymptotic analysis shows that there are no interior orbits that
end at $\bar{\varphi}=0$ and $v=+\infty$).

\subsection{Fixed points of the dynamical system}

Some of the fixed points of the dynamical system~\eqref{Dynsysuv}
depend on $\lambda$ and thereby on the positive potential.
Although not necessary, apart from the asymptotic condition~\eqref{lambda.lim.0}
we, for simplicity, assume that the potentials also satisfy the second
condition~\eqref{lambda.lim.infty} and that the potentials are
\begin{itemize}
\item[(i)] \emph{monotonically decreasing}, {\it i.e.} $V_{,\varphi} <0$ and hence $\lambda(\varphi)>0$, or
\item[(ii)] \emph{have a single positive minimum}, {\it i.e.} there exists a positive finite
$\varphi = \varphi_0$ such that $V_{,\varphi}|_{\varphi=\varphi_0}=0$ and
$V_{,\varphi\varphi}|_{\varphi=\varphi_0}>0$
and hence $\lambda(\varphi_0)=0$, $\lambda_{,\varphi}|_{\varphi=\varphi_0}<0$.\footnote{Incidentally,
as far as we know, asymptotically inverse power-law potentials with a positive
minimum have not been investigated before in the literature.}
\end{itemize}
%

\subsubsection*{The matter dominant boundary $v=0$}

It follows from~\eqref{Dynsysuv} that the boundary set $v=0$ is independent of
$\lambda$ and thereby also on the potential and that $v=0$ contains three
Friedmann-Lema\^{i}tre lines of fixed points:
%
\begin{subequations}\label{FL.fixed}
\begin{alignat}{2}
\mathrm{FL}_0^{\varphi_*}&\!: &\qquad (\bar\varphi,u,v) &= (\bar{\varphi}_*,0,0),\\
\mathrm{FL}_{\pm}^{\varphi_*}&\!: &\qquad (\bar\varphi,u,v) &= (\bar{\varphi}_*,\pm\sqrt2,0),
\end{alignat}
\end{subequations}
where the constant $\bar{\varphi}_*$ satisfies $0\leq\bar{\varphi}_*\leq1$. Note that
$w_\varphi=-1$ on $\mathrm{FL}_0^{\varphi_*}$ while $w_\varphi=1$ on
$\mathrm{FL}_{\pm}^{\varphi_*}$. These fixed points correspond to the fixed points
with the same labels in AUW~\cite{alhetal23} (see equations (24)).
The lines of fixed points are connected by heteroclinic orbits (a heteroclinic orbit 
is a solution trajectory that connects two different fixed points)
$\mathrm{FL}_\pm^{\varphi_*} \rightarrow\mathrm{FL}_0^{\varphi_*}$ that are straight lines
with `frozen' scalar field values $\bar{\varphi} = \bar{\varphi}_* = \mathrm{constant}$.
The orbit structure on the boundary $v=0$ is shown in Figure~\ref{Fig1a}.

\subsubsection*{The matter dominant boundary $\bar{\varphi}=0$ \label{tracker}}

The boundary set $\bar{\varphi}=0$ reflects
the unboundedness of $\lambda$, and this gives rise to new fixed points with $v>0$,
as follows from~\eqref{Dynsysuv}:
\begin{subequations} \label{fixed.points}
\begin{align}
\mathrm{T}:\quad (\bar{\varphi},u,v) &= u_\mathrm{T}\left(0,1,\frac{1}{pb}\right),\qquad u_\mathrm{T} = \sqrt{\frac{p}{2+p}},\\
\mathrm{B}:\quad (\bar{\varphi},u,v) &= \sqrt{2}\left(0,-1, \frac{1}{4b}\right),
\end{align}
\end{subequations}
where $b = F(0)>0$ is the constant defined in~\eqref{def.F}.
\emph{On the boundary set} $\bar{\varphi}=0$, a local analysis shows that the fixed point
$\mathrm{T}$ is a spiral sink,\footnote{The eigenvalues associated with $\mathrm{T}$ on
the boundary $\bar{\varphi}=0$ are rather complicated but can conveniently be described
as follows: $\lambda_1 + \lambda_2 =  -\frac32\left(\frac{p + 4}{p + 2}\right) - 3u_\mathrm{T}b$,
$\lambda_1\lambda_2 = \frac92(p + 4)u_\mathrm{T}b$.}
$\mathrm{B}$ is a source (although $\mathrm{B}$ is a saddle in the full state space)
and the three FL fixed points with $v=0$ are saddles. We conjecture that all
interior orbits in the boundary set $\bar{\varphi}=0$ are attracted to $\mathrm{T}$.
Although the details of the orbit structure on $\bar{\varphi}=0$ depend on the
parameters $p$ and $b$, there are no bifurcations as the parameters change and thus
the qualitative orbit structure on $\bar{\varphi}=0$ is independent of the potential;
a representative description is shown in Figure~\ref{Fig1b} using the values
$b=1$ and $p=3$.
%

\subsubsection*{The exponential boundary $\bar{\varphi}=1$}

We have already noted that this boundary set coincides with
the boundary set $\bar{\varphi}=1$ in AUW~\cite{alhetal23}.
We briefly summarize the fixed points: apart from the $\bar{\varphi}=1$ boundary
end points of the matter-dominated FL lines of
fixed points with $v=0$ there are additional
fixed points on the exponential $\bar{\varphi} = 1$ boundary:
the scalar field dominated `kinaton' fixed points $\mathrm{K}_\pm$,
the de Sitter fixed point $\mathrm{dS}$ (when $\lambda_+=0$),
the power-law fixed point $\mathrm{P}$ (when $0 < |\lambda_+| < \sqrt{6}$),
and the scaling fixed point $\mathrm{S}$ (when $|\lambda_+| > \sqrt{3}$).
The values of $u$ and $v$ for the fixed points on the $\bar{\varphi}=1$
boundary are:
\begin{subequations} \label{fixed.points.exp}
\begin{align}
\mathrm{K}_\pm\!:\quad (u,v) &= \left(\pm\sqrt{2},\sfrac{1}{\sqrt{3}}\right),\\
\mathrm {dS}:\quad (u,v) &= \left(0,\sfrac{1}{\sqrt{3}}\right),\qquad\,\, \lambda_+=0,\\
\mathrm{P}\!:\quad (u,v) &= \sfrac{1}{\sqrt{3}}(\lambda_+,1),
\qquad 0 < |\lambda_+| < \sqrt{6},\label{Pfixedpoint}\\
\mathrm{S}\!:\quad (u,v) &= \left(\text{sgn}(\lambda_+),\sfrac{1}{|\lambda_+|}\right), \qquad |\lambda_+| > \sqrt{3}.
\end{align}
\end{subequations}
Details and figures depicting the orbit structures for the different cases associated
with $\lambda_+ =0$, $0<|\lambda_+|<\sqrt{3}$, $\sqrt{3}<|\lambda_+|<\sqrt{6}$, $\sqrt{6}<|\lambda_+|$
on the exponential boundary $\bar{\varphi}=1$ are given in AUW~\cite{alhetal23} while global results
were proven in~\cite{alhetal22}. For monotonically decreasing potentials $\mathrm{P}$
is a sink (as is $\mathrm{dS}$ with $q=-1$ when $\lambda_+=0$); since
$q = -1 + \lambda_+^2/2$ at $\mathrm{P}$ it follows that $\lambda_+ < \sqrt{2}$ results
in future eternal acceleration, which we, for simplicity, henceforth assume
when the potential is monotonically decreasing.
%
%

\subsubsection*{The scalar field dominant boundary $v\bar{\varphi}=1/\sqrt3$\label{dSo}}

For monotonically decreasing scalar field potentials with
$\lambda(\varphi)>0$ and $0 \leq \lambda_+ < \sqrt{2}$, the function $3H^2$
in~\eqref{3H2mon} is also a monotonic function on the interior of this boundary,
although the evolution of $3H^2$ goes through an inflection point for orbits when
and if they pass through $u=0$, which they only can do once since
$u^\prime|_{u=0}>0$, and hence there are no periodic orbits or
fixed points in the interior of the scalar field dominant boundary in this case.

\begin{figure}[ht!]
	\begin{center}
			\subfigure[$-\sqrt{6} < \lambda_+ <0$]{\label{Fig2a}
			\includegraphics[width=0.4\textwidth]{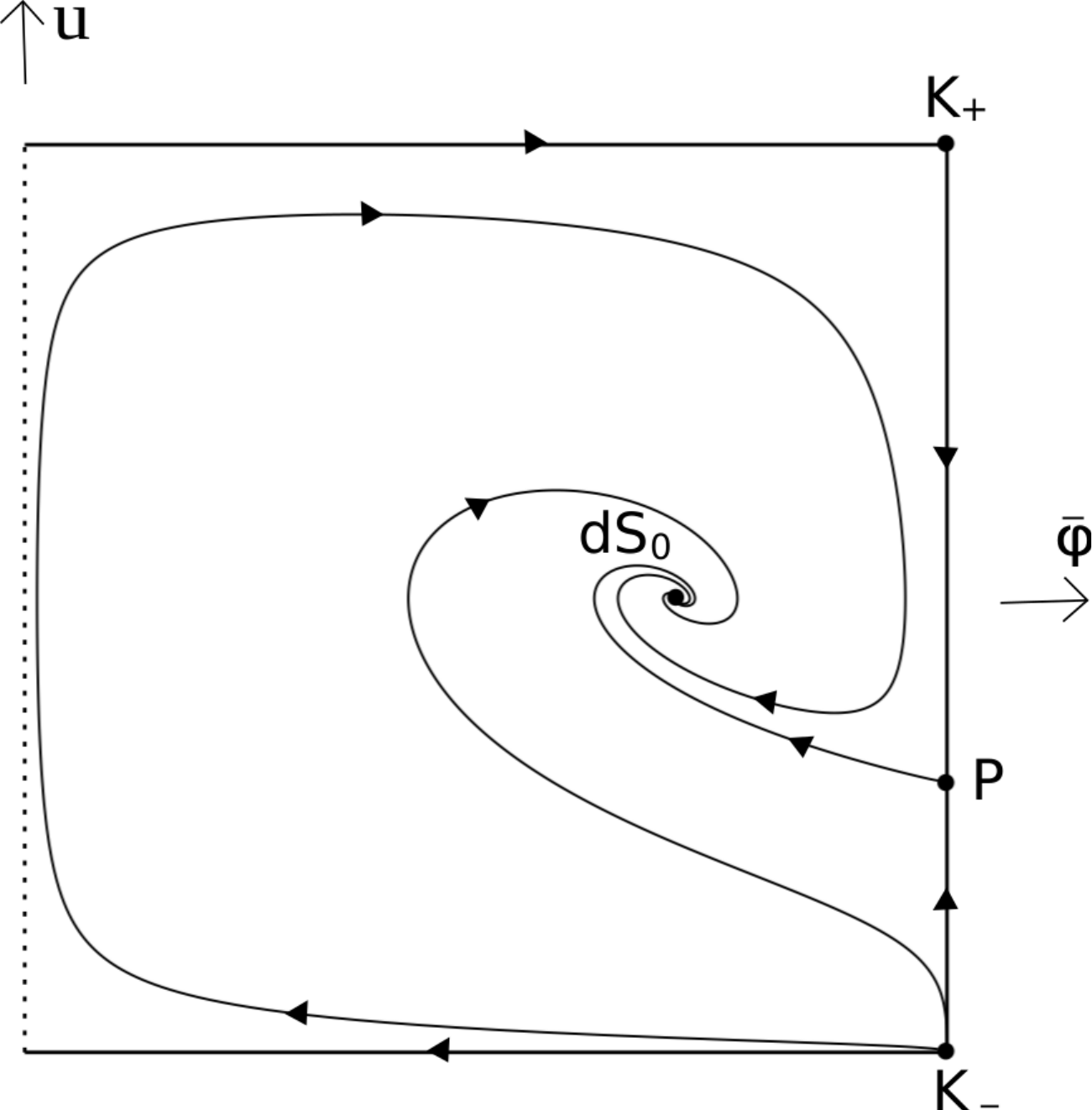}}
		\hspace{0.5cm}
			\subfigure[$\lambda_+ \leq -\sqrt{6}$]{\label{Fig2b}
	\includegraphics[width=0.4\textwidth]{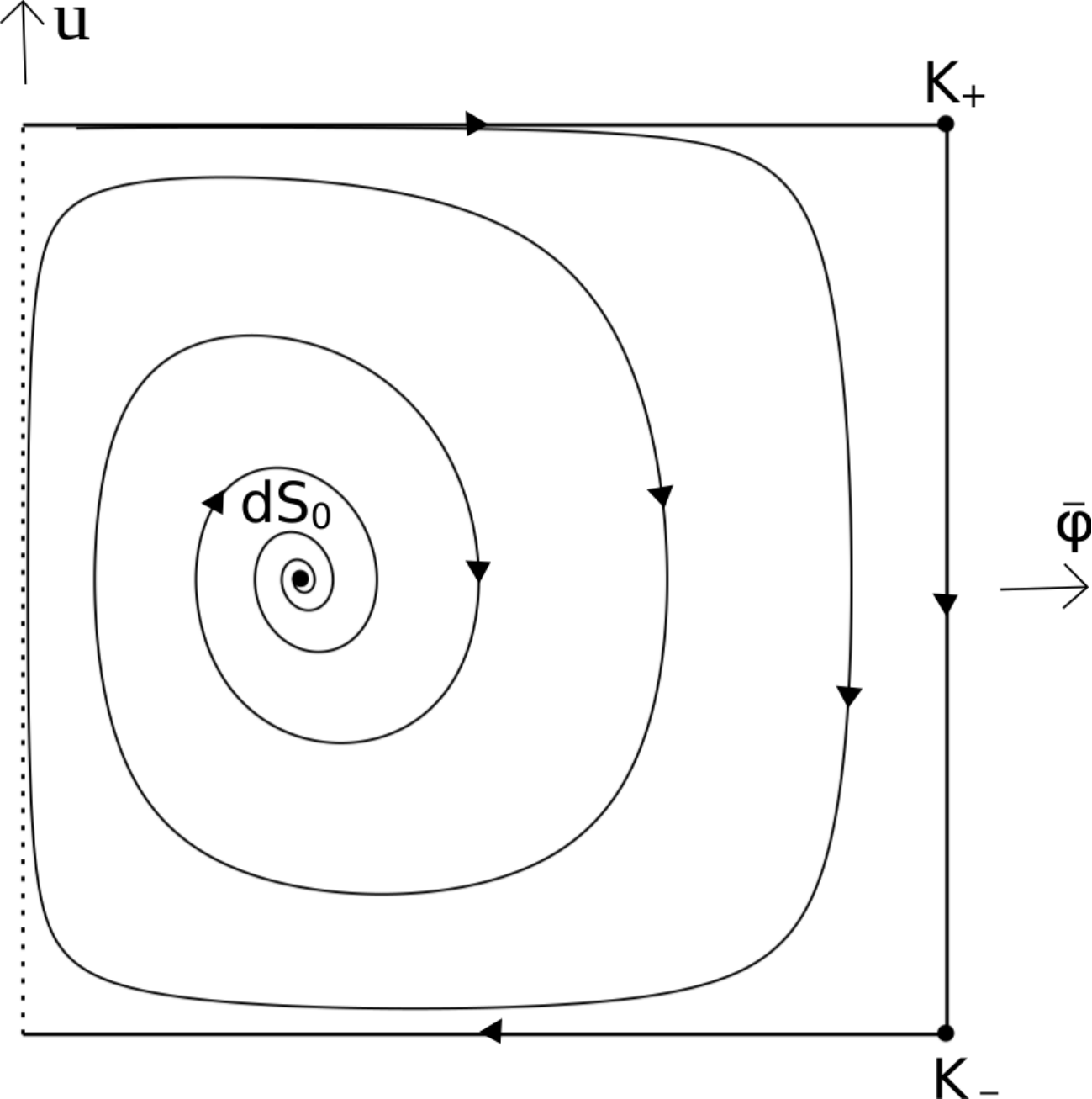}}
		\vspace{-0.5cm}
	\end{center}
	\caption{The orbit structure on the scalar field dominant
             boundary $v\bar{\varphi}=1/\sqrt{3}$ projected onto
             $(\bar{\varphi},u)$ with $0<\bar\varphi\leq1$ on the horizontal axis and
             $u\in [-\sqrt2, \sqrt2]$ on the vertical axis for the cases
             $-\sqrt{6} < \lambda_+ <0$ and $\lambda_+ \leq -\sqrt{6}$, respectively.
             These two cases are illustrated by the models
             given in eq.~\eqref{lambda.hyp.gen} with $p=1/2$, $\nu=2$ and the two values $\alpha = -1$ and
             $\alpha=-10$, which result in $\lambda_+=-1$ and $\lambda_+ = - 10$, respectively,
             since $\lambda_+ = p\nu\alpha$ for these models.
             The dotted line corresponds to the cut off in $\bar{\varphi}$ on the
             scalar field dominant boundary in Figure~\ref{Fig1}.}
            \label{Figminbound}
\end{figure}

Let us now turn to the case of a scalar field potential with a positive minimum.
Apart from the previous fixed points, the minimum results in an additional
fixed point given by
\begin{equation} \label{DS0.fixed}
\mathrm{dS}^{0}\!:\quad (\bar\varphi, u,v)
= \left(\bar\varphi_0, 0,\sfrac{1}{\sqrt3\,\bar{\varphi}_0}\right),
\end{equation}
where $\lambda(\bar\varphi_0)=0$, $\bar{\varphi}_0 \in (0,1)$, at the minimum of
the potential. Since
$\lambda_{,\varphi}|_{\bar{\varphi} = \bar{\varphi}_0} = -(V_{,\varphi\varphi}/V)|_{\bar{\varphi} = \bar{\varphi}_0}$
it follows that $\lambda_{,\varphi}|_{\bar{\varphi} = \bar{\varphi}_0} < 0$ yields a
positive minimum of the potential, and when this is the case, which we assume,
the eigenvalues associated with $\mathrm{dS}^{0}$ have negative real parts
and therefore $\mathrm{dS}^{0}$ is then a sink,\footnote{The
eigenvalues of the fixed point $\mathrm{dS}^{0}$ are
$-(3/2)\left(1\pm\sqrt{1+(4/3)\lambda_{,\varphi}|_{\bar{\varphi} = \bar{\varphi}_0}}\right)$, $-3$,
where the eigenvectors of the first pair are tangential to the scalar field
dominant boundary while the eigenvector connected with the eigenvalue $-3$ corresponds to the
$\Lambda$CDM orbit associated with the positive minimum of the scalar field potential $V$.\label{eigenvalues.dS0} }
which is a spiral on the scalar field dominant boundary if
$\lambda_{,\varphi}|_{\bar{\varphi} = \bar{\varphi}_0} < -3/4$.
When $-\sqrt6<\lambda_+< 0$ the fixed point $\mathrm{P}$ exists, with one orbit originating
from it ($\mathrm{P}$ is replaced with $\mathrm{dS}$ if $\lambda_+=0$) on the scalar field
dominant boundary, but this fixed point leaves the state space when $\lambda_+ = -\sqrt{6}$,
which results in that all orbits on the scalar field dominant boundary (apart from the
fixed point $\mathrm{dS}^0$) originate from an asymptotic cycle of boundary orbits and the
limit $v\rightarrow +\infty$ where $u$ is monotonically increasing when
$\lambda_+\leq-\sqrt{6}$, see Figure~\ref{Figminbound}.\footnote{Heuristically the situation is quite
similar to when $\lambda$ is bounded and $\lim_{\varphi\rightarrow - \infty}\lambda = \lambda_-\geq \sqrt{6}$,
$\lim_{\varphi\rightarrow + \infty}\lambda = \lambda_+ \leq -\sqrt{6}$, as
discussed in AUW~\cite{alhetal23}; in both cases the `scalar field particle' is increasing its energy
toward the past since there is friction toward the future, and in both cases the scalar field particle
bounces infinitely many times between two steep potential walls toward the past, where the present potential
wall at $\varphi=0$ is even steeper than the potential wall in AUW~\cite{alhetal23} corresponding to
$\lambda_- \geq \sqrt{6}$. It is therefore not surprising that in appropriate variables one can describe
this phenomenon as a heteroclinic limit cycle, {\it i.e} a cyclic chain of heteroclinic orbits.}


%

We conclude this section by first summarizing the values of $\Omega_\varphi$ and $w_\varphi$
at the fixed points:
\begin{subequations}\label{fixed.points.om.w}
\begin{alignat}{3}
\Omega_\varphi &= 1\!:\qquad \mathrm{P},\,\mathrm{dS},\,\mathrm{dS}^{0},\,
\mathrm{K}_{\pm}\qquad &\text{with}\qquad
w_\varphi &= \sfrac{\lambda_+^2}{3} - 1,-1,-1,1 ,\\
\Omega_\varphi &=0\!:\qquad \mathrm{FL}_0^{\varphi_*},\, \mathrm{FL}_\pm^{\varphi_*},\,
\mathrm{T},\, \mathrm{B}\qquad &\text{with}\qquad w_\varphi &= -1,1, -\sfrac{2}{2+p},1,\\
\Omega_\varphi &= \sfrac{3}{\lambda_+^2}\!:\qquad \mathrm{S}\qquad &\text{with}\qquad
w_\varphi &= 0.
\end{alignat}
\end{subequations}
As follows from~\eqref{q.expr}, the deceleration parameter is given by
$q = (1+3w_\varphi\Omega_\varphi)/2$ and can be read off from the above values.


Finally, we notice that for the class of potentials under
consideration the behaviour at late times of future accelerating models ($q<0$)
is of three possible types:
\begin{itemize}
\item[i)] Monotonically decreasing potentials: if $0< \lambda_+ <\sqrt2$
then the fixed point $\mathrm P$ is the sink, with $w_\varphi=\sfrac{\lambda_+^2}{3} - 1$
and $q = \sfrac{\lambda_+^2}{2} - 1$.
\item[ii)] Monotonically decreasing potentials: if $\lambda_+=0$ then the fixed point
$\mathrm{dS}$ is the sink, with $w_\varphi=-1$ and $q=-1$.
\item[iii)] Potentials with a positive minimum: the fixed point $\mathrm{dS}^0$ is the
sink, with $w_\varphi=-1$ and $q=-1$.
\end{itemize}
We conjecture, supported by heuristical arguments and numerical experiments, that in each case
the sink is the \emph{future global attractor}, denoted by ${\cal A}$, which thereby
attracts all interior orbits, including those relevant for quintessence.

\section{Quintessence\label{sec:cosmconsequences}}

In this section we use the new dynamical system to describe and
compare thawing, scaling freezing and tracking quintessence.

\subsection{Observationally viable quintessence models}

The first step is to identify which orbits in the ski-slope state
space $(\bar{\varphi},u,v)$ that describe observationally viable quintessence models.
We begin with two necessary conditions:
\begin{itemize}
\item The model must be \emph{accelerating
at present and late times}, {\it i.e.} $q<0,$ where $q$ is given by~\eqref{q.expr}.
\item The model must have an \emph{early long matter-dominated epoch}
($\Omega_\varphi\lesssim 0.03$ for $\Delta N \sim 10$, see AUW~\cite{alhetal22})
and satisfy $\Omega_\varphi\approx 0.68$ at
the present time $N=0$.
\end{itemize}
%

First we identify the orbits in the state space that begin
at matter-dominated fixed points and then end at the future
attractor ${\cal A}$. Then we regard these unstable manifolds as reference
orbits in the sense that \emph{initial data sufficiently close to
a fixed point of a viable reference orbit also provide viable quintessence models.}
Referring to~\eqref{fixed.points.om.w} we see that
the relevant fixed points  are $\mathrm{FL}_0^{\varphi_*}$ with
$0<\bar{\varphi}_*<1$, $\mathrm{T}$, and also $\mathrm{S}$ if the potential
has a positive minimum and $\lambda_+ \ll -1$.
For a given potential belonging to the present class of potentials with unbounded
$\lambda(\varphi)$ there are thereby three possibilities for reference orbits for
generating viable quintessence models, which we denote symbolically as follows:
\begin{itemize}
\item[i)] a one parameter family of reference orbits
$\mathrm{FL}_0^{\varphi_*}\rightarrow{\cal A}$
with $0<\bar{\varphi}_*<1$,
\item[ii)] a single reference orbit $\mathrm{S}\rightarrow{\cal A}=\mathrm{dS}^0$
when $\lambda_+ \ll -1$,
\item[iii)] a single reference orbit $\mathrm{T}\rightarrow{\cal A}$,
\end{itemize}
where the future attractor ${\cal A}$ for monotonically decreasing potentials
is given by $\mathrm{dS}$, $\mathrm{P}$ when $\lambda_+ =0$,
$0 < \lambda_+ < \sqrt{2}$, respectively, while ${\cal A} = \mathrm{dS}^0$
when the potential has a positive minimum. The reference orbits in all three cases
are the unstable manifolds of the fixed points from which they originate.
As we will see, the three sets of fixed points $\mathrm{FL}_0^{\varphi_*}$,
$\mathrm{S}$ and $\mathrm{T}$ have stable manifolds that push nearby orbits
(corresponding to open sets of initial data near the fixed points) to
the unstable manifold orbits which they subsequently shadow until they all end
at the future attractor ${\cal A}$. Next we discuss the three cases,
which correspond to thawing, scaling freezing and tracking quintessence,
respectively, in more detail.

\subsubsection*{Thawing quintessence}

In case i) each fixed point of $\mathrm{FL}^{\varphi^*}_0$ has one
eigenvalue that is zero (corresponding to the line of fixed points), one that
is positive, which yields one reference orbit
$\mathrm{FL}_0^{\varphi_*}\rightarrow{\cal A}$ for each fixed point
of $\mathrm{FL}_0^{\varphi_*}$ (the unstable manifold of
$\mathrm{FL}_0^{\varphi_*}$), and one negative eigenvalue corresponding to the
stable manifold $v=0$ of $\mathrm{FL}^{\varphi^*}_0$ on which $\bar{\varphi}$
is constant/`frozen', shown in Figure~\ref{Fig1}. The condition
of a long matter-dominated epoch is \emph{very} restrictive and leads to initial data with
$0<v\ll 1$, which result in open sets of orbits that are attracted/pushed toward
to the unstable manifold reference orbits of $\mathrm{FL}^{\varphi^*}_0$, which
they subsequently very closely shadow, until they all end at the future attractor
${\cal A}$. Note further that the equation of state for the reference orbits
starts at $w_\varphi\approx -1$ ($w_\varphi = -1$ at $\mathrm{FL}^{\varphi^*}_0$)
and then increases, which identify the reference orbits (and the orbits that shadow them)
as \emph{thawing quintessence} solutions.\footnote{See,
for example Chiba {\it et al}. (2013)~\cite{chietal13}, the introduction.}
We discussed and elaborated on this type of quintessence in a dynamical systems
setting for potentials with bounded ${\lambda}$ in AUW~\cite{alhetal23} and
notice that the situation for thawing quintessence is quite similar in the present
case with unbounded $\lambda$.

\subsubsection*{Scaling freezing quintessence}

Case ii) occurs for a potential with a positive minimum. A
matter-dominated scaling epoch for such potentials, where $\rho_\varphi$
approximately scales in time as $\rho_\mathrm{m}$, only holds for an open set of
interior initial data sufficiently close to the fixed point $\mathrm{S}$. Moreover, since
matter-dominance requires $\Omega_\varphi\lesssim 0.03 \Rightarrow v \lesssim 0.1$
and since $v|_\mathrm{S} = 1/|\lambda_+|$ it follows that matter-dominated
scaling orbits require $\lambda_+ \lesssim - 10$. The fixed point $\mathrm{S}$
has two negative eigenvalues with the boundary $\bar{\varphi}=1$ as the
stable manifold on which $\mathrm{S}$ attracts all orbits,
and one positive eigenvalue, which makes $\mathrm{S}$ a saddle point in the full
state space, where the reference orbit $\mathrm{S}\rightarrow\mathrm{dS}^0$ is
the unstable manifold associated with the positive eigenvalue. The open set of
interior matter-dominated scaling initial data sufficiently close to $\mathrm{S}$ first
approach $\mathrm{S}$ and then the reference orbit, which they then shadow,
and then they all finally end at $\mathrm{dS}^0$. Due to that these orbits, and
the reference orbit, exhibit the scaling property during the matter-dominated epoch
near $\mathrm{S}$ where $w_\varphi \approx w_\mathrm{m} =0$, and since $w_\varphi = -1$
at $\mathrm{dS}^0$, these orbits correspond to \emph{scaling freezing quintessence}
solutions, see {\it e.g.} Tsujikawa (2013)~\cite{tsu13}.

\subsubsection*{Tracking quintessence}

In this paper we focus on case iii).
The fixed point $\mathrm{T}$ has two eigenvalues with negative real parts with 
the matter dominant boundary $\bar{\varphi}=0$ as its stable manifold on which
$\mathrm{T}$ attracts all orbits and one positive eigenvalue, making $\mathrm{T}$
a saddle point in the full state space where the positive eigenvalue yields
the unstable manifold reference orbit $\mathrm{T}\rightarrow{\cal A}$.

Since, according to~\eqref{DeltaG},
$\Delta = \bar{\varphi}\lambda(\bar{\varphi})\,v/u = G(\bar{\varphi})v/u$
where $G(0) = p\,b$, $u_\mathrm{T} = p\,b\,v_\mathrm{T}$, it follows that
$\Delta=1$ \emph{at} the fixed point $\mathrm{T}$, {\it i.e.} the first tracker
condition is exactly fulfilled at $\mathrm{T}$. Since we can write
$\lambda=(p/\varphi )f(\varphi)$, where $f$ is regular and satisfies
$f(0)=1$ due to the boundedness condition~\eqref{lambda.lim.0}
and since $\Gamma = 1-\lambda_{,\varphi}/\lambda^2$ it follows that
$\Gamma-1 = 1/p > 1$ when $\varphi \rightarrow 0$ and hence at $\mathrm{T}$,
while $w_\varphi|_\mathrm{T}= u_\mathrm{T}^2 - 1 = -2/(2+p)<0$ and thus the second tracker
condition is also fulfilled at $\mathrm{T}$. We therefore refer to $\mathrm{T}$
as the \emph{tracker fixed point} and the unstable manifold reference orbit of
$\mathrm{T}$ as the \emph{tracker orbit}. When $w_\varphi(N)$ is nearly constant,
which is true near $\mathrm{T}$,\footnote{As we will see in section~\ref{sec:graphs},
there is a special class of models for which $w_\varphi = \mathrm{constant}$ holds
for the entire tracker orbit, but this is a very special class of models.}
the graph of $w_\varphi(N)$ has a horizontal plateau
$w_\varphi\approx-2/(2+p)$, as illustrated in the graphs of $w_\varphi(N)$
in section~\ref{sec:graphs}.

If $\bar{\varphi}$ is sufficiently close to the matter dominant boundary
$\bar{\varphi}=0$ this results in an open set of initial conditions that yields orbits
with a long matter-dominated epoch where they a) shadow the orbits on $\bar{\varphi} =0$,
see Figure~\ref{Fig1b}, b) then approach and spiral around $\mathrm{T}$,
where they still are matter-dominated but where being close to $\mathrm{T}$
also implies that the orbits approximately obey the tracker
conditions,\footnote{Note, however, that the scalar field effectively is a test field
not affecting the space-time geometry during matter-dominance.}
c) and where they subsequently track/shadow the tracker orbit until they all end
at the future attractor ${\cal A}$. Steinhardt {\it et al.} (1999)~\cite{steetal99}
characterizes the open set of orbits that track/shadow the tracker orbit as
undershooting and overshooting orbits. In terms of our state space description this
characterization is based on solutions with initial data close to $\bar{\varphi}=0$
with $u\lesssim0$, $0\lesssim v<v_\mathrm{T}$ (undershooting solutions), and
$u\gtrsim 0$, $v\gg v_\mathrm{T}$ (overshooting solutions).
We will give specific examples in section~\ref{sec:graphs}.

In this way, to quote
Steinhardt {\it et al}. (1999)~\cite{steetal99} (see the introduction),
who dealt with monotonically decreasing potentials
for which $\lambda_+=0$, primarily the inverse power-law potential:
"...a very wide range of initial conditions rapidly converge to a common, cosmic
evolutionary track." However, it is worth noticing that for monotonically decreasing
potentials with ${\cal A} = \mathrm{dS}$ the fixed point $\mathrm{dS}$ has an
eigenvalue that is zero with a stable interior state space center manifold direction,
and two eigenvalues with negative real parts with the boundary $\bar{\varphi}=1$ as 
their associated stable manifold. As a consequence \emph{all} orbits are strongly attracted 
to the center manifold of $\mathrm{dS}$, which means that they \emph{all} `track' each other,
including tracking/shadowing the tracker orbit.

Finally, we comment on the dynamical systems approach by Bahamonde {\it et al.} (2018)
used in~\cite{bahetal18} to study tracking quintessence for the inverse power-law potential.
To do so they introduced variables that result in a non-regular system that breaks down
when $\varphi\rightarrow 0$ (when their variable $z$ is one, see their eqs. 4.37-4.38).
They then regularize their equations by changing the time variable to obtain the
system 4.40-4.42. However, this hides that the new time variable asymptotically
breaks down when $\varphi\rightarrow 0$ and that their variables deform and crush our
regular dynamical system at $\bar{\varphi}=0$. This results in half circles at $z=1$ and
a fixed point at the intersection of two lines of fixed points at $x=0$ on the line of
fixed points $B_x$, see Figure 9 in~\cite{bahetal18}. The eigenvalues at this fixed point
are all zero, see Table 7 in~\cite{bahetal18}. The authors then perform a numerical
investigation that suggests that the tracker orbit originates from the fixed point $x=0$ on
$B_x$.

In contrast to our regularized dynamical system with a time variable that does not break
down when $\varphi\rightarrow 0$, the variables and dynamical system 4.40-4.42
in~\cite{bahetal18} cannot be used to: 1) explain the tracking attractor feature, which
our variables show is due to that $\bar{\varphi}=0$ is the stable manifold of $\mathrm{T}$,
2) nor can the dynamical system 4.40-4.42 be used to obtain analytic
approximations for the tracker orbit since all eigenvalues are zero at $x=0$ on  $B_x$,
while we will use the unstable manifold corresponding to the positive eigenvalue at $\mathrm{T}$
to obtain simple and accurate approximations for tracker orbits for a wide range of
potentials in a future follow up paper.

\subsection{Thawing, scaling freezing and tracking quintessence}

We conclude this section with some further comments on thawing, scaling
freezing and tracking quintessence for a potential with unbounded $\lambda$.
We first note that there is a close mathematical relationship between the present models with
unbounded $\lambda(\varphi)$ and tracking quintessence and models with bounded $\lambda(\varphi)$
that have a very large $\lambda_-$ and thereby admit scaling quintessence associated with
$\varphi\rightarrow-\infty$. The two cases have a fixed point $\mathrm{T}$ and $\mathrm{S}$,
respectively, with stable manifolds on a boundary associated with their respective scalar field
limits, and both have a single reference orbit as their unstable manifold that attracts an open
set of orbits that come close to the fixed point; cf. AUW~\cite{alhetal23} with the present
paper, especially the illustrative figures in AUW and in the next section. However, recall that
$\Omega_\varphi = 3/\lambda_+^2$, $1 \ll |\lambda_+|$ and hence $0<\Omega_\varphi \ll 1$ at
$\mathrm{S}$ for scaling freezing quintessence while $\Omega_\varphi = 0$ for
tracking (and thawing) quintessence at $\mathrm{T}$ (and $\mathrm{FL}^{\varphi_*}$).

We finally comment on the distinction between tracking, scaling freezing and thawing
quintessence, which is not as clear cut as it first appears. We first note that scalar field
potentials \emph{always} give rise to an open set of thawing quintessence solutions (where
some of the thawing solutions are observationally compatible with the $\Lambda$CDM model if
a scalar field potential has a sufficiently slowly changing part)
shadowing the thawing quintessence reference orbits.
If the present class of potentials exhibits a positive minimum and
$\lambda_+\ll - 1$ then apart from the open sets of orbits
that exhibit thawing and tracking quintessence there also exists
an open set of scaling freezing quintessence orbits.
The coexistence of thawing quintessence with tracking and also scaling freezing quintessence
(if the potential has a minimum and $\lambda_+ \ll - 1$)
causes complications as regards distinguishing thawing quintessence from
tracking and scaling freezing quintessence. This complication is caused
by the orbits $\mathrm{FL}_0\rightarrow\mathrm{T}$ and $\mathrm{FL}_0\rightarrow\mathrm{S}$
in the boundary sets $\bar\varphi=0$ and $\bar\varphi=1$, respectively.
It follows that the unstable manifold of any fixed point
$\mathrm{FL}_0^{\varphi_*}$ with $0< \bar{\varphi}_*\ll 1$ ($0 \ll \bar{\varphi}_* < 1$) will shadow
the orbit $\mathrm{FL}_0\rightarrow\mathrm{T}$ ($\mathrm{FL}_0\rightarrow\mathrm{S}$)
and hence pass close to $\mathrm{T}$ ($\mathrm{S}$), giving rise to solutions that are both
thawing and tracking (scaling freezing) quintessence models. To avoid this ambiguity we impose the
restriction $0\ll \bar{\varphi}_*<1$ (and $0 \ll \bar{\varphi}_* \ll 1$ for potentials with
a minimum and $\lambda_+ \ll - 1$) when describing thawing quintessence models.
\section{Numerical simulations for tracking quintessence \label{sec:graphs}}

Instead of using the inverse power-law potential as an illustrative example,
we use the following more versatile class of hyperbolic potentials:
\begin{equation}\label{lambda.hyp.gen}
V = V_0\left[\frac{\nu (\cosh\nu\varphi)^{1-\alpha}}{\sinh(\nu\varphi)}\right]^p
\quad\implies\quad \lambda = \frac{p\nu }{\tanh{\nu\varphi}}\left[1 + (\alpha-1)(\tanh{\nu\varphi})^2\right],
\end{equation}
where $\varphi \in (0,\infty)$, $p>0$, $\nu>0$, while $\alpha$ is arbitrary, and
\begin{equation}\label{lpillustrate}
\lambda_+ = p\nu\alpha.
\end{equation}
For these models a suitable choice of $\bar{\varphi}$ is
\begin{equation} \label{bar.varphi.def}
\bar\varphi = \tanh(\nu\varphi) \quad \implies \quad
F(\bar{\varphi})=\nu(1-\bar{\varphi}^2),
\end{equation}
satisfying~\eqref{def.F} with $b=\nu$,
%
%
which leads to
%
%
%
%
%
\begin{equation} \label{Gamma.hyp}
G(\bar{\varphi}) = p\nu\left[1 + (\alpha-1)\bar{\varphi}^2\right],\qquad
\Gamma - 1 = \frac{(1 - \bar{\varphi}^2)(1 - (\alpha-1)\bar{\varphi}^2)}{p[1 + (\alpha-1)\bar{\varphi}^2]^2},
\end{equation}
where $\Gamma > 1$ globally if $\alpha<2$. The special cases $\alpha=1$ and
$\alpha=0$ with $V \propto \sinh^{-p}(\nu\varphi)$ and
$V \propto \tanh^{-p}(\nu\varphi)$, respectively,
are well known.\footnote{See Urena-Lopez and Matos (2000)~\cite{uremat00} and
Bag {\it et al}. (2018)~\cite{bagetal18}.}

We note that if we use the illustrative positive
potential~\eqref{lambda.hyp.gen} a minimum of such a potential implies
that $\alpha<0$ and where the fixed point $\mathrm{dS}^0$ is located at
$\bar\varphi_0 = 1/\sqrt{1-\alpha}$.
In order for $\mathrm{dS}^0$ not only to be a sink but a spiral sink
it follows that $p\nu^2\alpha < - 3/8$, which follows from the spiral condition
$\lambda_{,\varphi}|_{\bar{\varphi} = \bar{\varphi}_0} < -3/4$ in
footnote~\ref{eigenvalues.dS0} where
$\lambda_{,\varphi}|_{\bar\varphi = \bar\varphi_0} = 2p\nu^2\alpha$ 
for the present models.
We will use these models to illustrate that there are three types of tracker orbits
for potentials that are monotonically decreasing: those for which $w_\varphi$ is
overall decreasing, constant, and increasing, and then we will
turn to the case of a potential with a positive minimum.

\subsection{Monotonic potentials: Three types of tracker orbits}

The central features for tracking quintessence are the tracker fixed point $\mathrm{T}$
and the tracker orbit $\mathrm{T}\rightarrow {\cal A}$, where ${\cal A}$
is one of the fixed points $\mathrm{P}$, $\mathrm{dS}$ for monotonically decreasing
potentials depending on whether $0< \lambda_+ < \sqrt{2}$ or $\lambda_+=0$, respectively.
We then note that
\begin{equation}
1 + w_\varphi|_\mathrm{T} = u_\mathrm{T}^2 = \frac{p}{2 + p},\qquad
1 + w_\varphi|_{\mathrm{P}/\mathrm{dS}} = u_{\mathrm{P}/\mathrm{dS}}^2 = \frac13\lambda_+^2 = \frac13(p\nu\alpha)^2,
\end{equation}
where the last equality follows from~\eqref{lpillustrate}. Hence
the overall decrease/increase of $w_\varphi$ for the tracker orbit
$\mathrm{T}\rightarrow {\cal A}$ is determined by the values of $u_\mathrm{T}$
and $u_\mathrm{P}$ (or, possibly, $u_\mathrm{dS}$ when $w_\varphi$ is overall decreasing),
which leads to:
\begin{subequations}
\begin{alignat}{2}
\sqrt{3}u_\mathrm{T} &= \sqrt{\frac{3p}{2+p}} > \sqrt{3}u_{\mathrm{P}/\mathrm{dS}} = \lambda_+,&\qquad &\text{overall decrease in $w_\varphi$},\\
\sqrt{3}u_\mathrm{T} &= \sqrt{\frac{3p}{2+p}} = \sqrt{3}u_\mathrm{P} = \lambda_+,&\qquad &\text{overall constant $w_\varphi$},\label{w.tracker.2}\\
\sqrt{3}u_\mathrm{T} &= \sqrt{\frac{3p}{2+p}} < \sqrt{3}u_\mathrm{P} = \lambda_+,&\qquad &\text{overall increase in $w_\varphi$},
\end{alignat}
\end{subequations}
where we recall that $\lambda_+=p\nu\alpha$ for the potentials~\eqref{lambda.hyp.gen}.

It does not follow in general that $w_\varphi$ is constant
during the evolution for the borderline case~\eqref{w.tracker.2}, although there is no
overall change in $w_\varphi$ during the evolution from $\mathrm{T}$ to
$\mathrm{P}$. However, suitable restrictions on the parameters $\alpha$,
$p$, $\nu$ for the potentials~\eqref{lambda.hyp.gen} give rise to a subclass of
potentials for which $w_\varphi=u^2-1$ \emph{is} constant for a
particular solution. In the early years of quintessence,
Sahni and Starobinsky (2000)~\cite{sahsta00},
equation (121), and Urena-Lopez and Matos (2000)~\cite{uremat00}, investigated models
with the potential~\eqref{lambda.hyp.gen} with $\alpha=1$, {\it i.e.}
$V \propto \sinh^{-p}(\nu\varphi)$, and pointed out that these models admitted a
special solution for which $w_\varphi$ is constant. This model corresponds to the
tracker orbit when $w_\varphi|_\mathrm{T} = w_\varphi|_\mathrm{P}$ and hence
$\sqrt{3}u_\mathrm{T} = \sqrt{3}u_\mathrm{P} = \lambda_+ = p\nu$, where
the last equality follows from setting $\alpha=1$ in~\eqref{lpillustrate},
which results in $\sqrt{3p/(2+p)} = p\nu$, and hence
\begin{equation} \label{w.const}
\nu = \sqrt{\frac{3}{p(2+p)}}.
\end{equation}
Positive potentials $V \propto \sinh^{-p}(\nu\varphi)$ with any value of $p>0$ and
the above value of $\nu$ result in that the tracker orbit $\mathrm{T}\rightarrow\mathrm{P}$
is a straight line in the state space $(\bar{\varphi}, u,v)$, parallel to the
$\bar{\varphi}$ axis given by $u=u_\mathrm{T} = \sqrt{p/(2 + p)}$,
$v = v_\mathrm{T} = u_\mathrm{T}/p\nu =1/\sqrt3$,
and thereby with a constant $w_\varphi$ given by $1+w_\varphi = u_\mathrm{T}^2 = p/(2+p)$.
Furthermore, since $\Omega_\varphi = 3v^2\bar{\varphi}^2$ it follows
that $\Omega_\varphi= \bar{\varphi}^2$.

\begin{figure}[ht!]
	\begin{center}
			\subfigure[Tracker orbit for the
		potential~\eqref{lambda.hyp.gen} with $\alpha=0$, and $\nu=p=3$.]{\label{Fig3a}
			\includegraphics[width=0.3\textwidth]{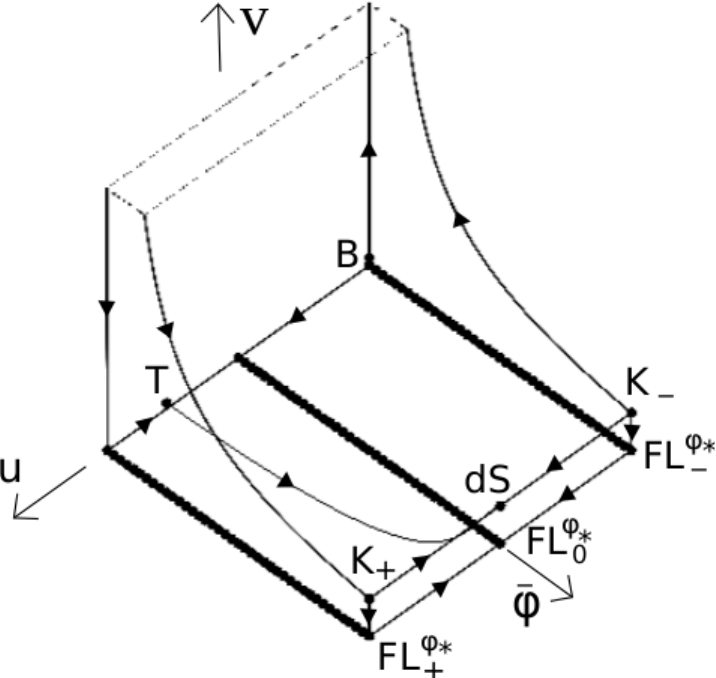}}
		\subfigure[Tracker orbit for the
		potential~\eqref{lambda.hyp.gen} with $\alpha=1$, $\nu=\sqrt{\frac{12}{5}}$ and $p=0.5$.]{\label{Fig3b}
			\includegraphics[width=0.3\textwidth]{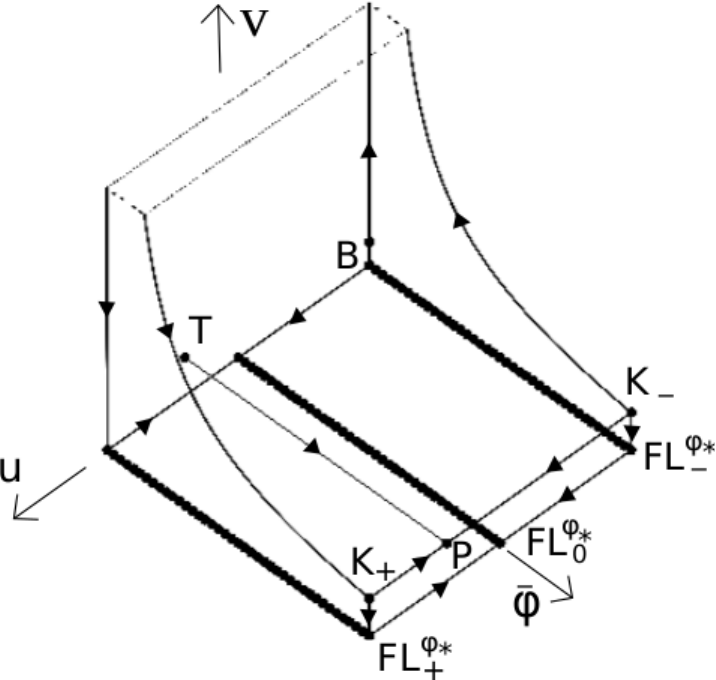}}
		\subfigure[Tracker orbit for the
		potential~\eqref{lambda.hyp.gen} with $\alpha=12$, $\nu=2$ and $p=0.05$.]{\label{Fig3c}
			\includegraphics[width=0.3\textwidth]{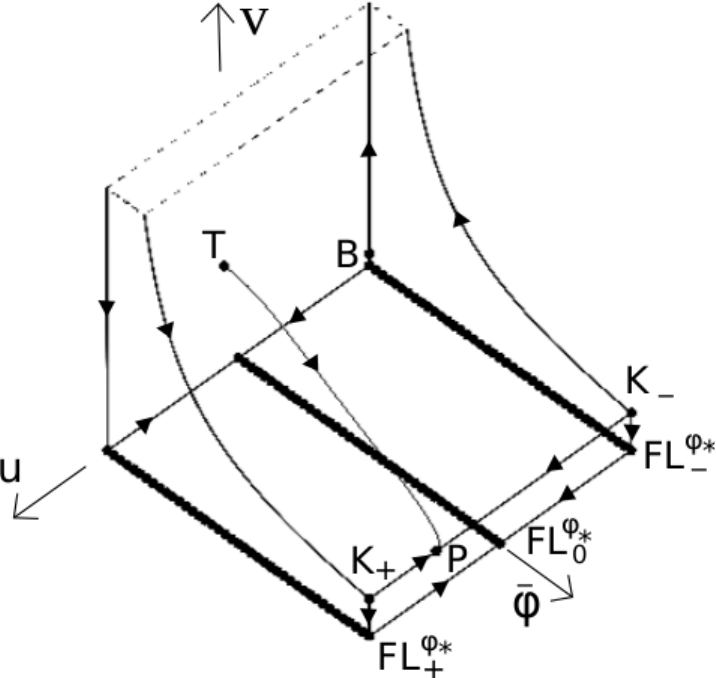}}	
		\subfigure[Decreasing $w_\varphi(N)$ for the tracker orbit with a potential~\eqref{lambda.hyp.gen}
                   with $\alpha=0$, and $\nu=p=3$.]{\label{Fig3d}
			\includegraphics[width=0.3\textwidth]{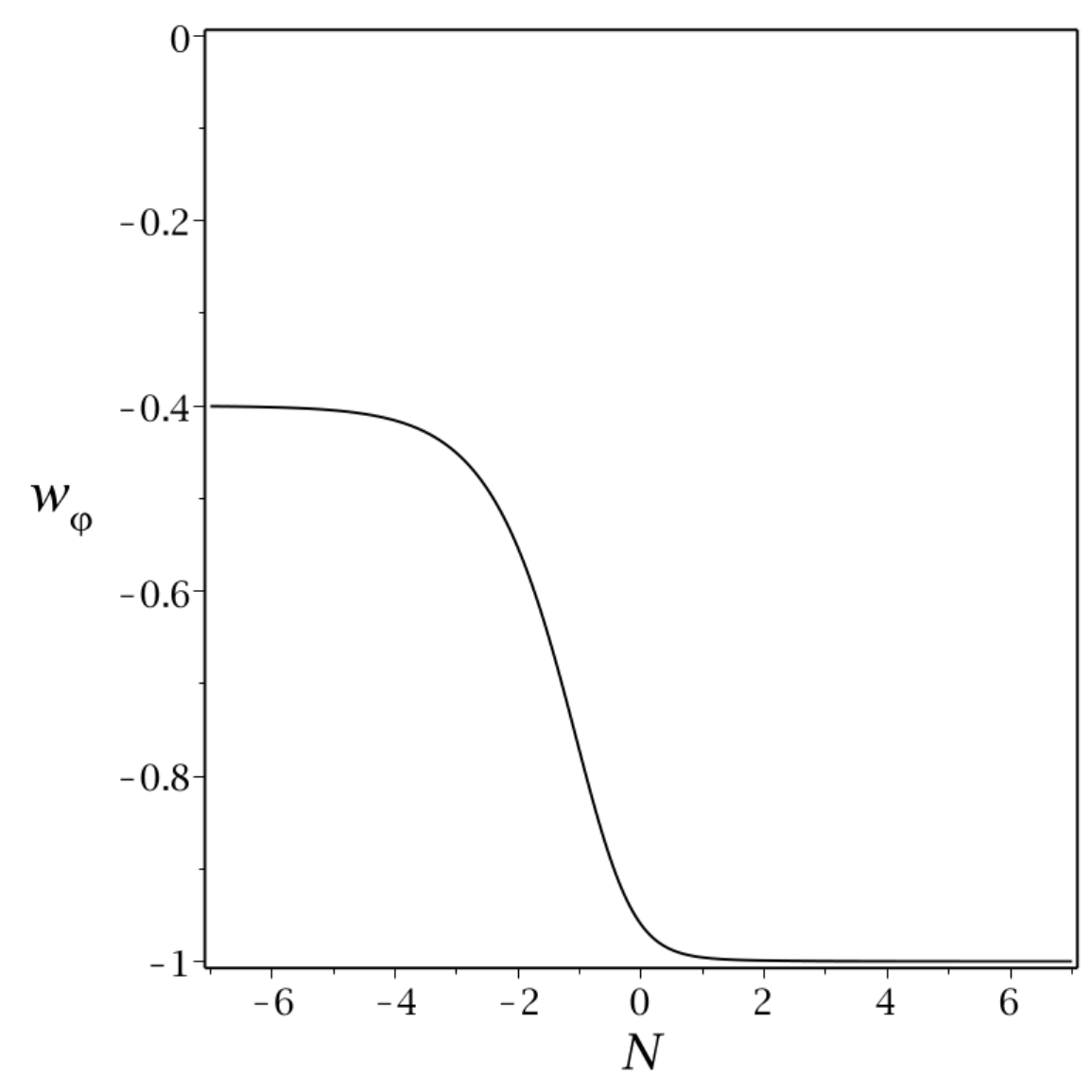}}
		\subfigure[Constant $w_\varphi(N)$ for the tracker orbit with a potential~\eqref{lambda.hyp.gen} with $\alpha=1$, $\nu=\sqrt{\frac{12}{5}}$ and
                   $p=0.5$.]{\label{Fig3e}
			\includegraphics[width=0.3\textwidth]{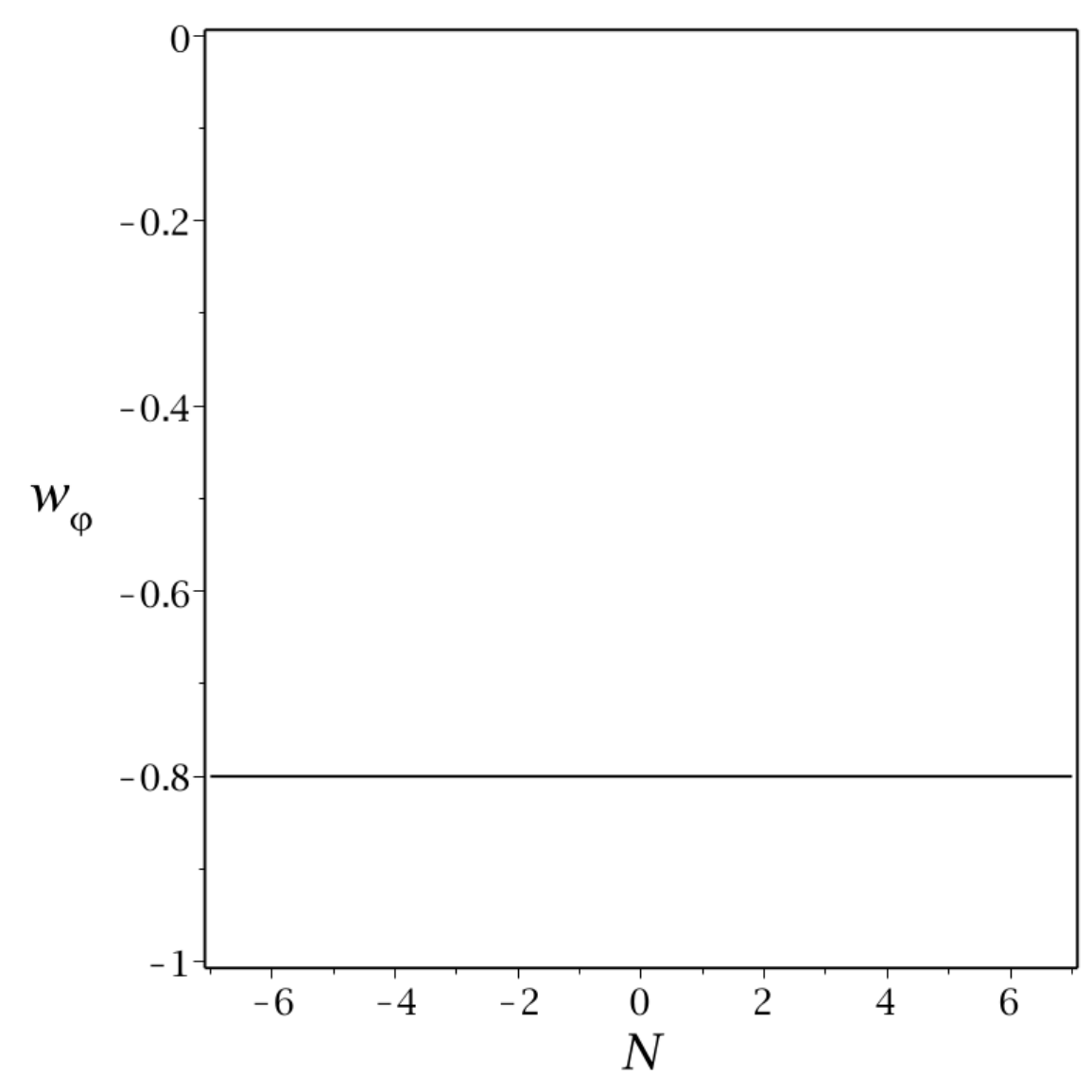}}
		\subfigure[Increasing $w_\varphi(N)$ for the tracker orbit with a potential~\eqref{lambda.hyp.gen} with $\alpha=12$, $\nu=2$ and
                   $p=0.05$.]{\label{Fig3f}
	\includegraphics[width=0.3\textwidth]{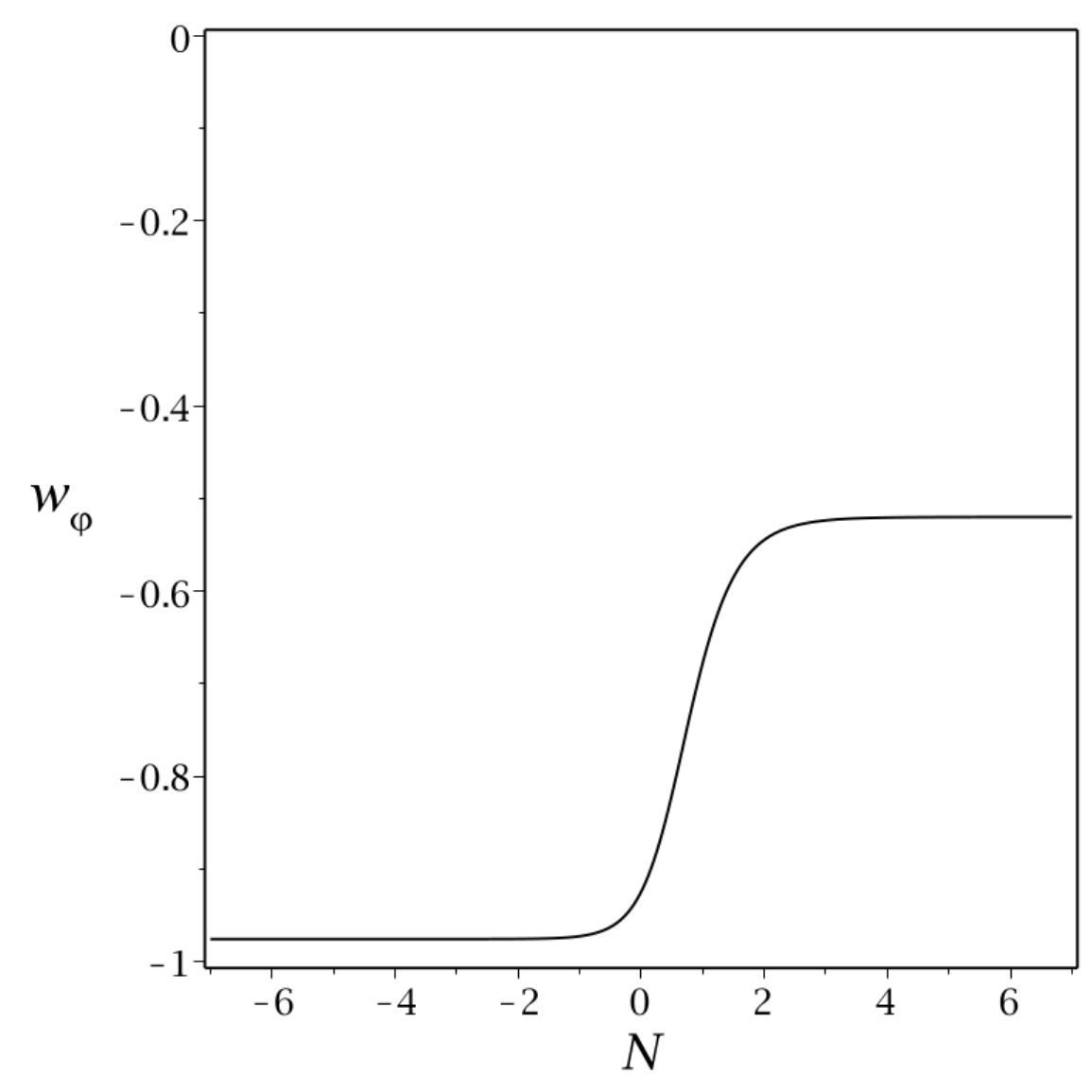}}	
		\vspace{-0.5cm}
	\end{center}
	\caption{Tracker orbits in the ski-slope state-space and the respective graph of $w_\varphi(N)$
        for the potential~\eqref{lambda.hyp.gen}, illustrating that $w_\varphi$ can decrease, be a constant,
        and increase for the tracker orbit.
}\label{Fig3}
\end{figure}

Along \emph{any} of the above three types of tracker
orbits, the graph of $w_\varphi$ begins during matter-dominance ($\Omega_\varphi\approx 0$)
close to $\mathrm{T}$ with a horizontal plateau $w_\varphi\approx -2/(2+p)$,
{\it i.e.} $-1<w_\varphi<0$. During the evolution $\Omega_\varphi$ thereby
starts close to zero near $\mathrm{T}$ and then increases 
to 1 at ${\cal A}$. The present epoch at $N=0$ is
characterized by that $\Omega_\varphi(N)$ reaches the observed value
$\Omega_{\varphi}(0) = \Omega_{\varphi,0}\approx 0.68$.
The value of $w_\varphi$ at late times depends on ${\cal A}$ but its value at
the present time depends on $\lambda(\bar{\varphi})$ at $\bar{\varphi}(N=0)$.
For monotonically decreasing potentials with $\lambda_+=0$ the tracker orbit
approaches $\mathrm{dS}$ and hence $w_\varphi \rightarrow - 1$ asymptotically,
but is greater than $-1$ when $N=0$. These features are illustrated in Figure~\ref{Fig3}.
%
%

\subsection{The tracker orbit for a potential with a minimum}

\begin{figure}[ht!]
	\begin{center}
		\subfigure[Tracker, $\Lambda$CDM, and a thawing quintessence orbit for the
		potential~\eqref{lambda.hyp.gen} with $\nu=2$, $p=1/2$, and  $\alpha=-1$.]{\label{Fig4a}
			\includegraphics[width=0.45\textwidth]{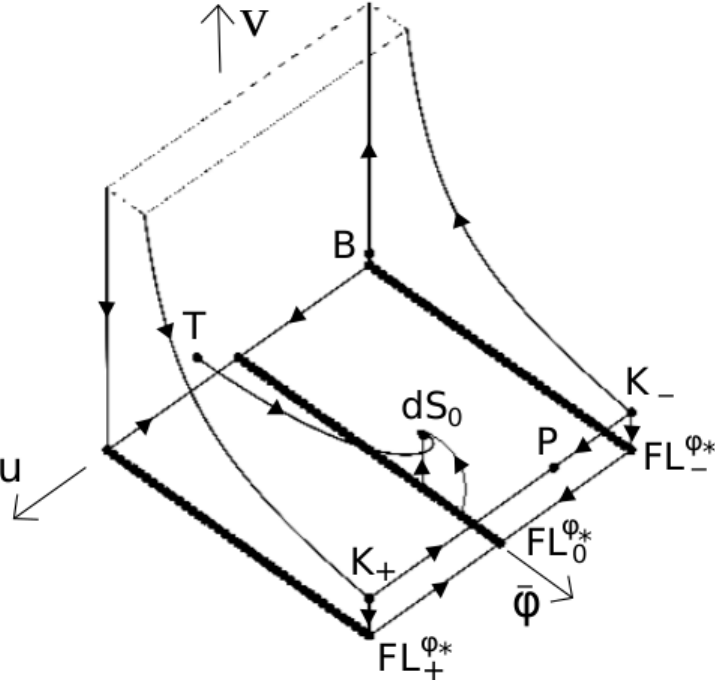}}	\hspace{1cm}
		\subfigure[Tracker, $\Lambda$CDM, scaling freezing, and a thawing quintessence orbit
          for the potential~\eqref{lambda.hyp.gen} with $\alpha=-10$, $\nu=2$ and $p=1/2$.]{\label{Fig4b}
			\includegraphics[width=0.45\textwidth]{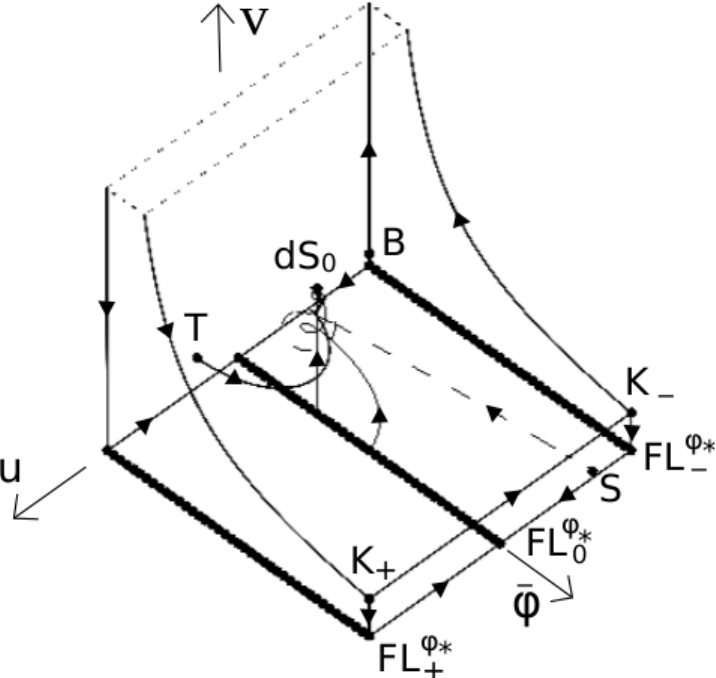}} \\
		\subfigure[Graphs of $w_\varphi$ for the tracker and thawing quintessence orbits for the
               potential~\eqref{lambda.hyp.gen} with $\nu=2$, $p=1/2$, and  $\alpha=-1$.]{\label{Fig4d}
			\includegraphics[width=0.4\textwidth]{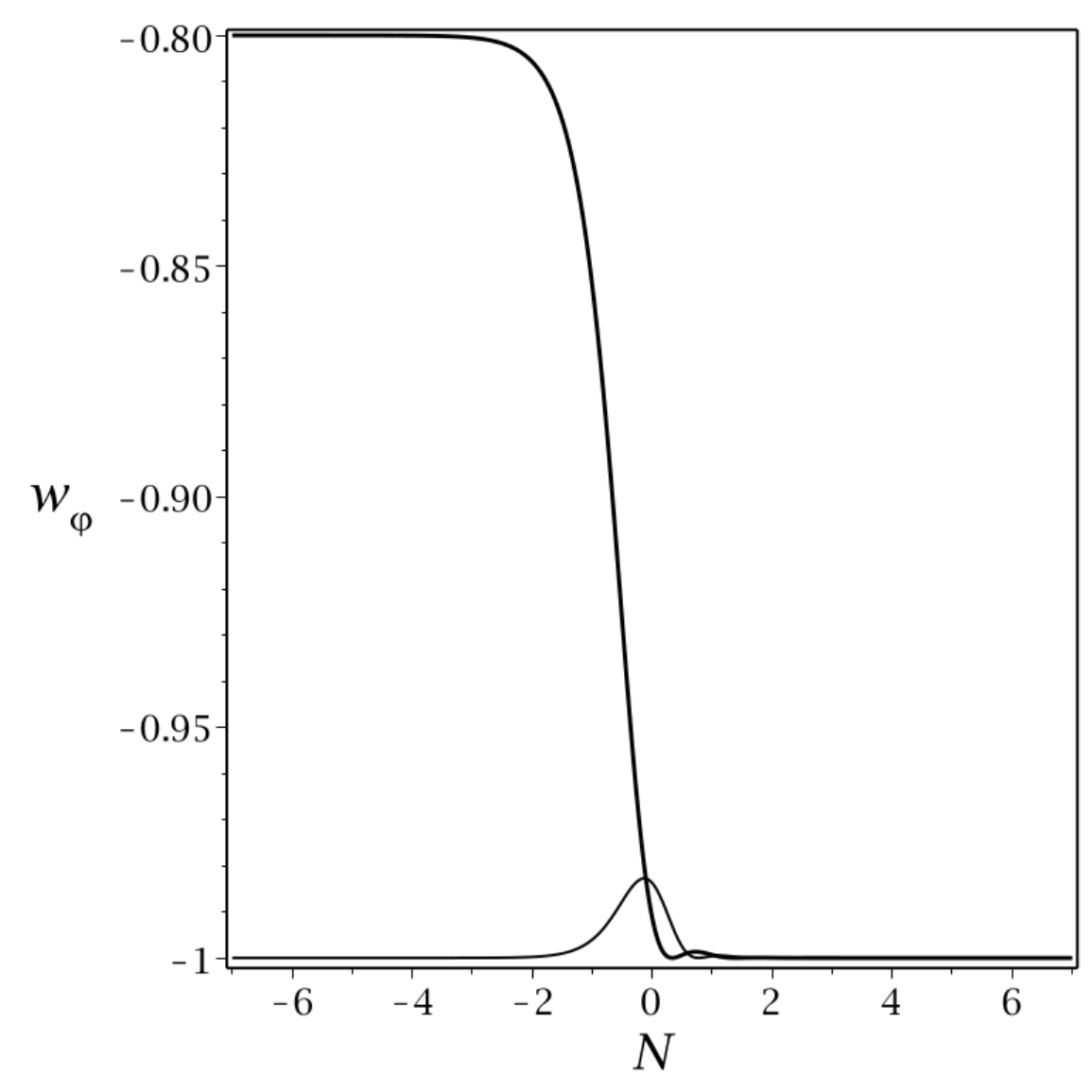}}	\hspace{1cm}
		\subfigure[Graphs of $w_\varphi$ for the tracker, thawing and scaling freezing quintessence orbits
             for the potential~\eqref{lambda.hyp.gen} with $\nu=2$, $p=1/2$, and  $\alpha=-10$.]{\label{Fig4e}
			\includegraphics[width=0.4\textwidth]{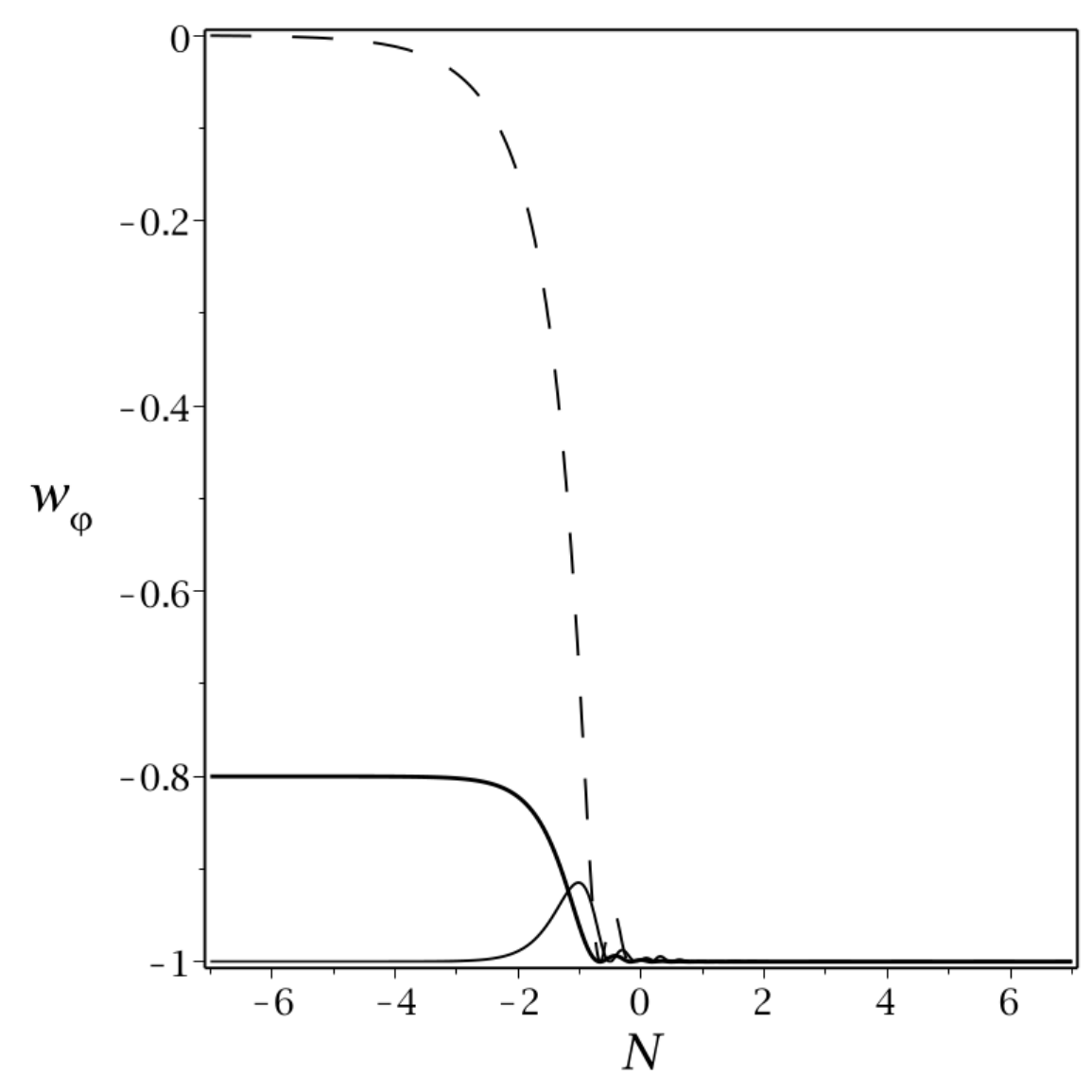}}
		\vspace{-0.5cm}
	\end{center}
	\caption{The tracker orbit in the ski-slope state-space and their graphs $w_\varphi(N)$ for the
		potential~\eqref{lambda.hyp.gen} with a minimum, together with the $\Lambda$CDM orbit, a thawing quintessence
        reference orbit originating from $\mathrm{FL}_0^{\varphi_*}$, and also a scaling freezing quintessence
        reference orbit when $\lambda_+ \ll -1$ in figure (b), with associated graphs of $w_\varphi$
        in (c) and (d), where $w_\varphi$ for the thawing orbit begins with
        $w_\varphi \approx -1$, while the tracker orbit begins with $w_\varphi \approx -2/(2+p)$,
        which for $p=1/2$ yields $w_\varphi \approx -0.8$, and where the dashed
        scaling freezing orbit begins with $w_\varphi \approx 0$.}\label{Fig4}
\end{figure}

In this case $\mathrm{T}\rightarrow\mathrm{dS}^0$ where the sink $\mathrm{dS}^0$
resides on the matter dominant boundary $v\bar{\varphi}=1/\sqrt3$ with
$\bar{\varphi}=\bar{\varphi}_0$ determined by $\lambda(\bar{\varphi}_0)=0$.
If $\lambda_{,\varphi}|_{\bar\varphi = \bar\varphi_0} <-3/4$,
then, according to footnote~\ref{eigenvalues.dS0}, $\mathrm{dS}^0$ is a spiral
focus sink in the boundary set $v\bar{\varphi}=1/\sqrt3$ (see Figures~\ref{Figminbound} and~\ref{Fig4}).
The fixed point $\mathrm{dS}^0$ has another feature which helps to describe how
the orbits are attracted to $\mathrm{dS}^0$, including the tracker orbit. There
is an orbit $\mathrm{FL}_0^{\varphi_*}\rightarrow \mathrm{dS}^0$ that is a
straight line characterized by $\bar{\varphi}=\bar\varphi_* = \bar{\varphi}_0$, $u=0$,
$0 < v < 1/\sqrt3$. This orbit describes the $\Lambda$CDM solution, which
corresponds to that the scalar field resides in the positive minimum of the potential,
where $\lambda(\bar{\varphi}_0) = 0 = G(\bar{\varphi}_0)$, giving rise to
$\Lambda= V(\bar{\varphi}_0)>0$. All interior orbits are asymptotic to
the spiral (assuming that
$\lambda_{,\varphi}|_{\bar\varphi = \bar\varphi_0} < -3/4$) sink $\mathrm{dS}^0$
and hence as they come close to
$\mathrm{dS}^0$ they spiral around the $\Lambda$CDM orbit. In particular,
the tracker orbit $\mathrm{T}\rightarrow\mathrm{dS}^0$ originates from ${\mathrm T}$
with $\Delta \approx 1$, $w_\varphi \approx \mathrm{constant}$ and then it
bends and eventually forms a spiral around the $\Lambda$CDM orbit as it approaches
$\mathrm{dS}^0$. These features are illustrated in Figure~\ref{Fig4}.
%

\subsection{Tracking quintessence for an open set of initial data}

%

\begin{figure}[ht!]
	\begin{center}
		\subfigure[Tracker, undershooting and overshooting orbits.]{\label{Fig5a}
			\includegraphics[width=0.30\textwidth]{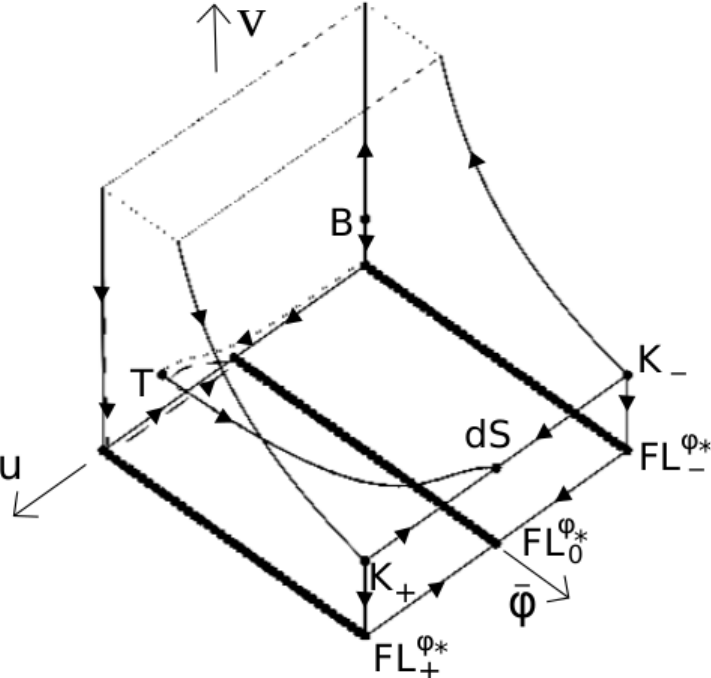}}	\hspace{0.3cm}
		\subfigure[Tracker, undershooting and overshooting orbits in the vicinity of $\bar{\varphi}=0$.]{\label{Fig5b}
			\includegraphics[width=0.30\textwidth]{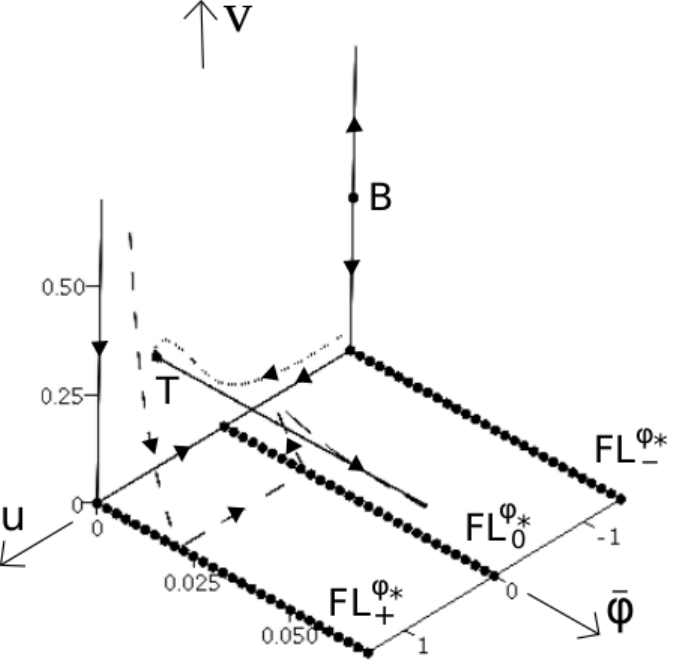}}\hspace{0.3cm}
		\subfigure[Graphs of $w_\varphi(N)$ for the tracker, undershooting and overshooting solutions.]{\label{Fig5c}
			\includegraphics[width=0.25\textwidth]{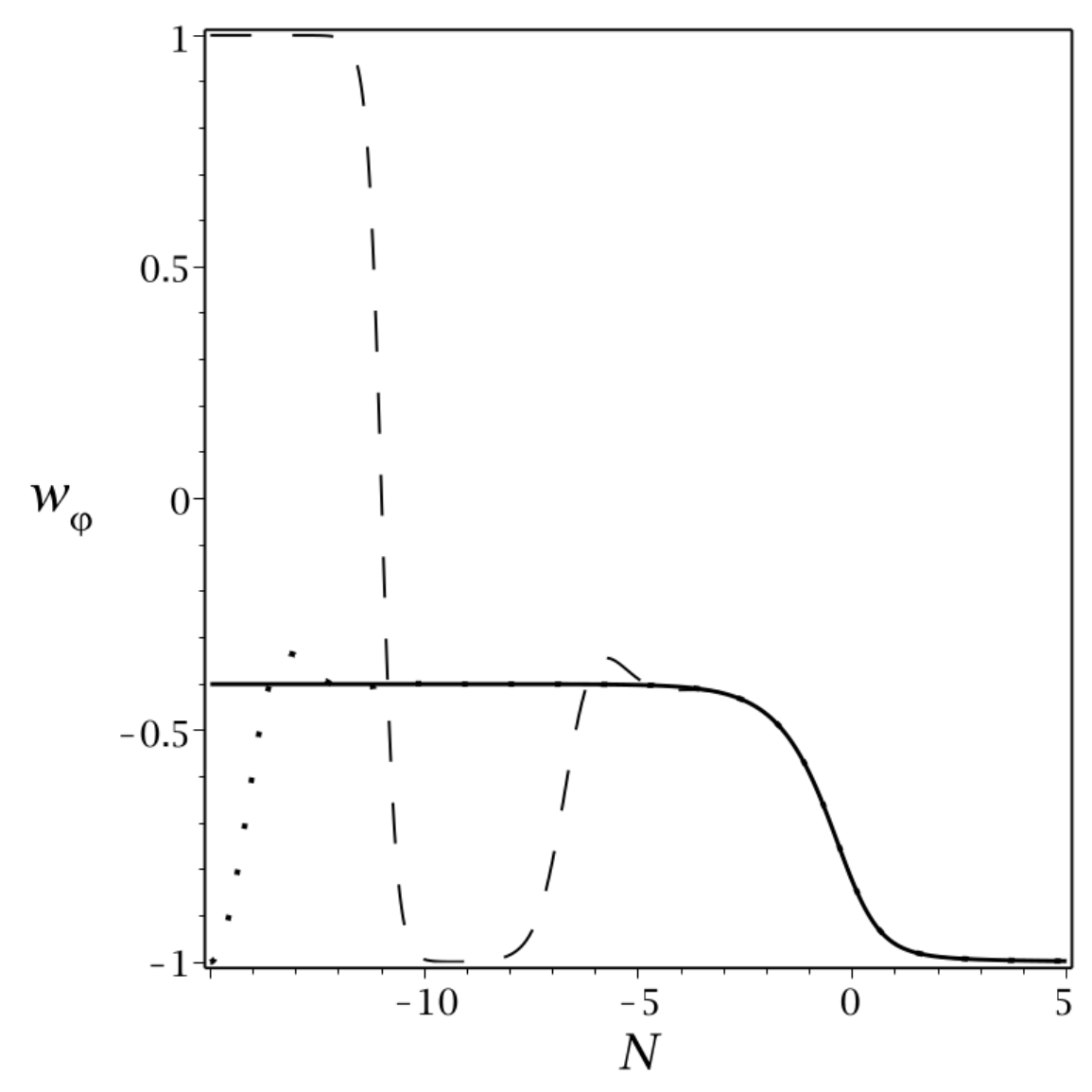}}
		\vspace{-0.5cm}
	\end{center}
	\caption{The tracker orbit (full line), overshooting (dashed line) and undershooting
    (dotted line) orbits in the ski-slope state-space and their graphs $w_\varphi(N)$ for
    the monotonically decreasing positive potential~\eqref{lambda.hyp.gen}
    with $\nu=1$, $p=3$ and $\alpha=0$.}\label{Fig5}
\end{figure}

Recall that Steinhardt {\it et al.} (1999)~\cite{steetal99} characterized
the open set of orbits tracking/shadowing the tracker orbit as undershooting and
overshooting orbits, which corresponds to
$u\lesssim0$, $0\lesssim v<v_\mathrm{T}$ (undershooting solutions) and
$u\gtrsim 0$, $v\gg v_\mathrm{T}$ (overshooting solutions), where the condition
of a long matter-dominated epoch requires these conditions for initial data
to be supplemented by a very small value of $\bar{\varphi}$ near the $\bar{\varphi}=0$
matter dominant boundary.\footnote{Note that $u<0$
results in $\bar{\varphi}'<0$, which means
that the boundary $\bar{\varphi}=0$ is stable (attracts nearby orbits),
while $u>0$ yields $\bar{\varphi}'>0$ and that $\bar{\varphi}=0$ is unstable.
The $u$-dependent stability properties of $\bar{\varphi}=0$ imply that
undershooting solutions begin tracking sooner than overshooting solutions,
see Figure~\ref{Fig5} and Figure 5 in Steinhardt {\it et al.} (1999)~\cite{steetal99}.}
Solutions corresponding to such initial data shadow
orbits on the $\bar{\varphi}=0$ boundary, where we can use
Figure~\ref{Fig1b} to obtain a feeling for how the solutions behave before they
begin to spiral around $\mathrm{T}$ in its vicinity during the matter-dominated
epoch. Since $w_\varphi = u^2 - 1$ this also gives a feeling
for how graphs of $w_\varphi$ behave, however, recall that the scalar field is essentially
a test field during the matter-dominated epoch and hence that this behaviour of
$w_\varphi$ is not an observable property and thereby physically unimportant.
These features are illustrated in Figure~\ref{Fig5}.
The graph $w_\varphi(N)$ in Figure~\ref{Fig5c} of the overshooting (dashed)
orbit in Figures~\ref{Fig5a} and~\ref{Fig5b}
covers the stage when it is very close to the boundary orbit at
$\bar{\varphi} = 0$, $u = \sqrt{2}$ where  $w_\varphi = 1$
(earlier it shadowed an orbit on the $\bar{\varphi}=0$ boundary that passed
$u=0$ at a very large value of $v$ not shown), subsequently, due to that
$\bar{\varphi}^\prime > 0$, it shadows a nearby orbit on the $u = \sqrt{2}$
boundary, see Figure~\ref{Fig5b}, which is followed by shadowing an
orbit $\mathrm{FL}_+^{\varphi_*} \rightarrow \mathrm{FL}_0^{\varphi_*}$
with $\bar{\varphi} = \mathrm{constant} \ll 1$, and then
an unstable $\mathrm{FL}_0^{\varphi_*}$ reference orbit near the
orbit $\mathrm{FL}_0^{\varphi_*}\rightarrow \mathrm{T}$ on
$\bar{\varphi} =0$,\footnote{This sequence of shadowing orbits is approximately
described by shadowing a certain orbit on the $\bar{\varphi}=0$ boundary, where
the approximation is improved by choosing a smaller $\bar{\varphi}$ datum.}
and finally it shadows the tracker orbit
$\mathrm{T}\rightarrow \mathrm{dS}$;\footnote{Note, as follows from
Figure~\ref{Fig5c}, the (dashed) overshooting solution only obeys the tracker conditions
for a short while between approximately $N=-5$ to $N=-3$. To increase this time period
requires an even smaller initial value of $\bar{\varphi}$ than presently, so that the orbit comes
closer to $\mathrm{T}$. Note also that this overshooting orbit illustrates the ambiguity between
thawing and tracker quintessence discussed in the previous section, since it shadows an unstable
$\mathrm{FL}_0^{\varphi_*}$ orbit during a thawing epoch.}
the graph $w_\varphi(N)$ in Figure~\ref{Fig5c} of the undershooting (dotted) orbit in
Figures~\ref{Fig5a} and~\ref{Fig5b} covers the stage when this undershooting orbit
shadows an orbit on the $\bar{\varphi}=0$ boundary when it passes $u=0$ where $w_\varphi = -1$
and afterwards where it, due to that $\bar{\varphi}^\prime <0$ when $u<0$, comes
extremely close to $\mathrm{T}$ and where it subsequently shadows the tracker orbit
$\mathrm{T}\rightarrow \mathrm{dS}$
(both the overshooting and undershooting orbits basically overlap with the tracker orbit
after the matter-dominated tracker stage, where the tracker solution therefore describes the
quintessence epoch of the overshooting and undershooting solutions extremely well).

%
%

\section{Concluding remarks\label{sec:concl}}

We have given, for the first time, a description of tracking quintessence in a
\emph{regular} state space framework using the $e$-fold time $N$,
showing that it is generated by any
potential for which $\lambda$ is unbounded with $\varphi\lambda\rightarrow p>0$ as
$\varphi\rightarrow0$. A central role is played by a unique tracker orbit that
originates from the tracker fixed point $\mathrm{T}$ which then governs the transition from
matter-domination to quintessence-domination and an accelerating universe.
Our state space provides some clarification for the claim of Steinhardt
{\it et al}. (1999)~\cite{steetal99} that with tracking quintessence,
"a very wide range of initial conditions rapidly converge to a common,
cosmic evolutionary track of $\rho_\varphi(N)$ and $w_\varphi(N)$".

We have shown that potentials with unbounded $\lambda$ also lead to thawing and
scaling freezing quintessence (the latter for potentials with a positive minimum
and $\lambda_+ \ll -1$) and have used the state space to make a
distinction between the three types, based on which fixed points the
reference orbits originate from (cf. with the more complicated classification
used in AUW\cite{alhetal23}). One might ask: \emph{Which of the open sets of
different types of quintessence initial data is preferred? Why should
tracking quintessence have a special status?}

Finally, we note that the present
dynamical systems formulation can be slightly modified to obtain simple
and accurate approximations with the $\Lambda$CDM model as a
continuous parameter/initial data limit for tracking and thawing quintessence,\footnote{This is
not possible for scaling freezing quintessence since $w_\varphi = w_\mathrm{m} = 0$
at $\mathrm{S}$ and not $w_\Lambda=-1$, although it is possible to obtain
approximate solutions for scaling freezing quintessence in the same manner as we will
accomplish for thawing and tracker quintessence.}
which will be the topic of a forthcoming paper.

\subsection*{Acknowledgments}
A. A. is supported by FCT/Portugal through CAMGSD, IST-ID, projects UIDB/04459/2020
and UIDP/04459/2020, and by the H2020-MSCA-2022-SE project EinsteinWaves, GA No. 101131233.
A.A. would also like to thank the CMA-UBI in Covilh\~a for kind hospitality.
C. U. would like to thank the CAMGSD, Instituto Superior T\'ecnico in Lisbon for kind hospitality.

\bibliographystyle{unsrt}
\bibliography{../Bibtex/cos_pert_papers}

\begin{thebibliography}{10}

\bibitem{rieetal98}
A.~G.~Riess et~al.
\newblock Observational evidence from supernovae for an accelerating universe
  and a cosmological constant.
\newblock {\em Astron. J.}, {\bf 116}:1009, 1998.

\bibitem{peretal99}
S.~Perlmutter et~al.
\newblock Measurements of omega and lambda from 42 high redshift supernovae.
\newblock {\em Astron. J.}, {\bf 517}:565, 1999.

\bibitem{caletal98}
R.~R. Caldwell, Rahul Dave, and Paul~J. Steinhardt.
\newblock Cosmological imprint of an energy component with general equation of
  state.
\newblock {\em Phys. Rev. Lett.}, {\bf 80}:1582--1585, 1998.

\bibitem{tsu13}
S.~Tsujikawa.
\newblock Quintessence: a review.
\newblock {\em Class. Quantum Grav.}, {\bf 30}:214003, 2013.

\bibitem{bahetal18}
S.~Bahamonde, C.~G. B{\"o}hmer, S.~Carloni, E.~J. Copeland, Wei Fang, and
  N.~Tamanini.
\newblock Dynamical systems applied to cosmology: Dark energy and modified
  gravity.
\newblock {\em Physics Reports}, {\bf 775-777}:1--122, 2018.

\bibitem{zlaetal99}
I.~Zlatev, L.~Wang, and P.~J. Steinhardt.
\newblock Quintessence, cosmic coincidence and the cosmological constant.
\newblock {\em Phys. Rev. Lett.}, {\bf 82}:896, 1999.

\bibitem{steetal99}
P.~J. Steinhardt, L.~Wang, and I.~Zlatev.
\newblock Cosmological tracking solutions.
\newblock {\em Phys.\ Rev.\ D}, {\bf 59}:123504, 1999.

\bibitem{peerat88}
P.~J.~E. Peebles and B.~Ratra.
\newblock Cosmology with a time variable cosmological constant.
\newblock {\em Astro. Phys. J.}, {\bf 325}:L17, 1988.

\bibitem{ratpee88}
B.~Ratra and P.~J.~E. Peebles.
\newblock Cosmological consequences of a rolling homogeneous scalar field.
\newblock {\em Phys. Rev. D}, {\bf 37}:3406, 1988.

\bibitem{RatQui92}
B.~Ratra and A.~Quillen.
\newblock {Gravitational lensing effects in a time-variable cosmological
  ‘constant’ cosmology}.
\newblock {\em Monthly Notices of the Royal Astronomical Society},
  259(4):738--742, 12 1992.

\bibitem{PodRat00}
S.~Podariu and B.~Ratra.
\newblock Supernova ia constraints on a time-variable cosmological
  “constant”.
\newblock {\em The Astrophysical Journal}, 532(1):109, mar 2000.

\bibitem{peerat03}
P.~J.~E. Peebles and Bharat Ratra.
\newblock The cosmological constant and dark energy.
\newblock {\em Rev. Mod. Phys.}, {\bf 75}:559--606, 2003.

\bibitem{alhetal23}
A.~Alho, C.~Uggla, and J.~Wainwright.
\newblock Quintessence from a state space perspective.
\newblock {\em Physics of the Dark Universe}, {\bf 39}:101146, 2023.

\bibitem{alhugg23}
A.~Alho and C.~Uggla.
\newblock Quintessential $\alpha$-attractor inflation: A dynamical systems
  analysis.
\newblock {\em Journal of Cosmology and Astroparticle Physics}, {\bf 11}:083,
  2023.

\bibitem{alhugg15b}
A.~Alho and C.~Uggla.
\newblock Scalar field deformations of lambda-cdm cosmology.
\newblock {\em Phys. Rev. D}, {\bf 92}(10):103502, 2015.

\bibitem{rubetal04}
C.~Rubano, P.~Scudellaro, and E.~Piedipalumbo.
\newblock Exponential potentials for tracker fields.
\newblock {\em Phys. Rev. D}, {\bf 69}:103510, 2004.

\bibitem{alhetal22}
A.~Alho, W.~C. Lim, and C.~Uggla.
\newblock Cosmological global dynamical systems analysis.
\newblock {\em Class. Quantum Grav.}, {\bf 39}:145010, 2022.

\bibitem{chietal13}
T.~Chiba, A.~De~Felice, and S.~Tsujikawa.
\newblock Observational constraints on quintessence: Thawing, tracker and
  scaling models.
\newblock {\em Phys. Rev. D}, {\bf 87}:083505, 2013.

\bibitem{uremat00}
L.A. Urena-Lopez and T.~Matos.
\newblock New cosmological tracker solution for quintessence.
\newblock {\em Phys. Rev. D}, {\bf 62}:081302, 2000.

\bibitem{bagetal18}
S.~Bag, S.S. Mishra, and V.~Sahni.
\newblock New tracker models of dark energy.
\newblock {\em Journal of Cosmology and Astroparticle Physics}, {\bf 08}:009,
  2018.

\bibitem{sahsta00}
V.~Sahni and A.Starobinsky.
\newblock The case for a positive cosmological lambda-term.
\newblock {\em Int. J. Mod. Phys. D}, {\bf 9}:373, 2000.

\end{thebibliography}

\end{document}